\documentclass{article}
\usepackage[T1]{fontenc}
\usepackage{amsmath,amssymb,amsthm}
\usepackage{comment}
\usepackage{mathtools}
\mathtoolsset{showonlyrefs}
\usepackage{dsfont}
\usepackage{mathrsfs}
\usepackage{enumitem}
\usepackage[a4paper, top=3cm, bottom=3.5cm, left=3.5cm, right=3.5cm]{geometry}
\usepackage[colorlinks,citecolor=blue,urlcolor=blue]{hyperref}
\hypersetup{pdftitle={Time-consistency of cash-subadditive risk measures: the infimal cocycle property}, pdfauthor={M. Ferrari, R. J. A. Laeven, E. Rosazza Gianin, M. Zullino}}
\usepackage[american]{babel}

\usepackage{csquotes}
\usepackage[
  backend=biber,
  style=alphabetic,     
  sorting=nyt,          
  maxnames=99,          
  maxalphanames=4,      
  minalphanames=3,      
  giveninits=false,     
  doi=true, eprint=true,
  url=false,            
  isbn=false
]{biblatex}
\appto\biburlsetup{\Urlmuskip=0mu\relax}
\appto{\bibsetup}{\emergencystretch=2em}
\renewbibmacro{in:}{%
  \ifentrytype{article}{}{\printtext{\bibstring{in}\intitlepunct}}}
\newcommand{\R}{\mathbb{R}}
\newcommand{\N}{\mathbb{N}}
\newcommand{\E}{\mathbb{E}}
\newcommand{\A}{\mathcal{A}}
\newcommand{\F}{\mathcal{F}}
\newcommand{\Q}{\mathbb{Q}}
\renewcommand{\P}{\mathbb{P}}
\newcommand{\1}{\mathds{1}}
\renewcommand{\d}{\,\mathrm{d}}
\newcommand{\ds}{\displaystyle}
\newcommand{\itemref}[1]{\textup{(\ref{#1})}}
\newcommand{\smallbullet}{{\mathbin{\vcenter{\hbox{\scriptsize$\bullet$}}}}}
\newcommand{\tinybullet}{{\mathbin{\vcenter{\hbox{\tiny$\bullet$}}}}}
\DeclareMathOperator*{\essmax}{ess\,max}
\DeclareMathOperator*{\essmin}{ess\,min}
 
\makeatletter
\newcommand{\proplabel}[2]{\phantomsection\protected@edef\@currentlabel{#1}\label{#2}}
\makeatother
 
\def\a{{\alpha}}
\def\r{{\rho}}
\def\l{{\lambda}}
\def\g{{\gamma}}
\def\eps{{\varepsilon}}
 
\DeclareMathOperator*{\esssup}{ess\,sup}
\DeclareMathOperator*{\essinf}{ess\,inf}
 
\numberwithin{equation}{section}
\newtheorem{theorem}{Theorem}[section]
\newtheorem{lemma}[theorem]{Lemma}
\newtheorem{proposition}[theorem]{Proposition}
\newtheorem{definition}[theorem]{Definition}
\theoremstyle{remark}
\newtheorem{remark}[theorem]{Remark}
\newtheorem{example}[theorem]{Example}

\title{Time-consistency of cash-subadditive risk measures:\\ The infimal cocycle property\thanks{
This research was funded in part by
the Netherlands Organization for Scientific Research (NWO) under grant NWO Vici 2020--2027 (Ferrari, Laeven, Zullino)
and by an UvA-AC3R Midsize Grant (Ferrari, Laeven).
Matteo Ferrari, Emanuela Rosazza Gianin, and Marco Zullino are members of Gruppo Nazionale per l’Analisi Matematica, la Probabilità e le loro Applicazioni (GNAMPA)-INdAM, Italy.
}}

\author{Matteo Ferrari \\
{\footnotesize Dept.~of Quantitative Economics}\\
{\footnotesize University of Amsterdam}\\
{\footnotesize and AC3R, The Netherlands}\\
{\footnotesize \texttt{M.Ferrari@uva.nl}}\\
\and Roger J.~A.~Laeven\footnote{Corresponding author.} \\
{\footnotesize Dept.~of Quantitative Economics}\\
{\footnotesize University of Amsterdam, AC3R, CentER}\\
{\footnotesize and EURANDOM, The Netherlands}\\
{\footnotesize \texttt{R.J.A.Laeven@uva.nl}}\\
\and Emanuela Rosazza Gianin \\
{\footnotesize Dept.~of Statistics and Quantitative Methods}\\
{\footnotesize University of Milano-Bicocca, Italy}\\
{\footnotesize \texttt{emanuela.rosazza1@unimib.it}}\\
\and Marco Zullino \\
{\footnotesize Dept.~of Mathematics and Applications}\\
{\footnotesize University of Milano-Bicocca, Italy}\\
{\footnotesize \texttt{M.Zullino@campus.unimib.it}}\\
}

\date{November 30, 2026}

\begin{document}

\maketitle

\begin{abstract}
    We characterize the time-consistency of convex, cash-subadditive dynamic risk measures in terms of their minimal penalty.
    The dual variables are conditional sub-probability densities, which factorize across an intermediate date in many ways, according to how the discount is allocated to the two periods.
    We introduce an \emph{infimal cocycle property}, in which the penalty is the infimum over all such factorizations, and show that it is equivalent to time-consistency, with the infimum attained.
    This holds on $L^\infty$ under continuity from below, and on $L^p$, $p<\infty$, without continuity assumptions.
    In the cash-additive case it reduces to the classical cocycle property.
    For risk measures induced by BSDEs, it represents the dynamic programming principle of a stochastic control problem whose value is the penalty.
\end{abstract}



\section{Introduction}
\label{sec:int}

Risk measures quantify the risk of financial positions, and are used to determine capital requirements  \cite{Artzner+Delbaen+Eber+Heath_1999_Coherent_measures_risk,Delbaen_2002_Coherent_risk_measures_general_probability_spaces,Follmer+Schied_2002_Convex_measures_risk_trading_constraints,Frittelli+RosazzaGianin_2002_Putting_order_risk_measures}.
In a dynamic setting, a risk measure $\r=(\r_{s,t})_{0\leq s\leq t}$ quantifies, at time $s$, the risk of a financial position $X$ with maturity $t$ by an $\F_s$-measurable random variable $\r_{s,t}(X)$ \cite{Delbaen_2006_Structure_m-stable_sets_particular_set_risk_neutral_measures,Riedel_2004_Dynamic_coherent_risk_measures,Frittelli+RosazzaGianin_2004_Dynamic_Convex_Risk_Measures,Detlefsen+Scandolo_2005_Conditional_dynamic_convex_risk_measures,Cheridito+Delbaen+Kupper_2006_Dynamic_monetary_risk_measures_bounded_discrete-time_processes}.
Most of the literature assumes \emph{cash-additivity}, $\r_{s,t}(X-k)=\r_{s,t}(X)+k$, which identifies cash available at time $s$ with cash available at the maturity $t$.
This is justified when positions are expressed in units of a known numéraire \cite{Farkas+Koch-Medina+Munari_2014_Beyond_cash-additive_risk_measures_changing_numeraire_fails}, but not under stochastic or ambiguous interest rates.
For this reason, El Karoui and Ravanelli \cite{ElKaroui+Ravanelli_2009_Cash_sub-additive_risk_measures_interest_rate_ambiguity} replaced cash-additivity with \emph{cash-subadditivity}, $\r_{s,t}(X-k)\leq\r_{s,t}(X)+k$ for $k\geq0$: cash withdrawn at maturity is discounted, and increases the risk at time $s$ by at most its amount.
Cash-subadditive risk measures have since been studied from several angles \cite{Cerreia-Vioglio+Maccheroni+Marinacci+Montrucchio_2011_Risk_measures_rationality_diversification,Frittelli+Maggis_2011_Dual_representation_quasi-convex_conditional_maps,Drapeau+Kupper_2013_Risk_preferences_robust_representation,Laeven+RosazzaGianin+Zullino_2023_Dynamic_return_star-shaped_risk_measures_BSDEs,Han+Wang+Wang+Xia_2025_Cash-subadditive_risk_measures_quasi-convexity}, and discounting ambiguity for cash flows is analyzed, e.g., in \cite{Acciaio+Follmer+Penner_2012_Risk_assessment_uncertain_cash_flows_discounting_ambiguity_bubbles,Penner+Reveillac_2015_Risk_measures_processes_BSDEs}, the latter also in connection with BSDEs.

In a dynamic setting, the central property of a risk measure is \emph{time-consistency},
\[
    \r_{s,t}(X)=\r_{s,u}\bigl(-\r_{u,t}(X)\bigr),\qquad 0\leq s\leq u\leq t,
\]
i.e., measuring the risk of $X$ directly at time $s$, or through an intermediate reevaluation at time $u$, gives the same result.
It makes the evaluations at different dates mutually coherent, and it underlies recursive valuation and the monitoring of risk over time;
see \cite{Acciaio+Penner_2011_Dynamic_risk_measures} and \cite[Chapter~11]{Follmer+Schied_2016_Stochastic_finance} for overviews.
In terms of acceptance sets, time-consistency is characterized by the decomposition $\A_{s,t}=\A_{s,u}+\A_{u,t}$ in the cash-additive case \cite{Cheridito+Delbaen+Kupper_2006_Dynamic_monetary_risk_measures_bounded_discrete-time_processes,Bion-Nadal_2008_Dynamic_risk_measures_time_consistency__risk_measures_BMO_martingales}, and by its leveled version
$\A^m_{s,t}=\bigcup_{n\in\A^m_{s,u}}\A^{-n}_{u,t}$
in general \cite{Ferrari+Laeven+RosazzaGianin+Zullino_2026_Characterization_risk_measures_BSDEs_acceptance_sets},  where $\A^m_{s,t}:=\{X\in L^p(\F_t):\r_{s,t}(X)\leq m\}$ and $\A_{s,t}:=\A^0_{s,t}$.
In terms of the dual representation's minimal penalty, for convex cash-additive risk measures time-consistency is equivalent to the \emph{cocycle property} $c_{s,t}(\Q)=c_{s,u}(\Q)+\E_\Q[c_{u,t}(\Q)|\F_s]$ for all probability measures $\Q\ll\P$ on $\F_t$ with $\Q=\P$ on $\F_s$ \cite{Bion-Nadal_2008_Dynamic_risk_measures_time_consistency__risk_measures_BMO_martingales,Bion-Nadal_2009_Time_consistent_dynamic_risk_processes,Follmer+Penner_2006_Convex_risk_measures_dynamics_their_penalty_functions,Delbaen+Peng+RosazzaGianin_2010_Representation_penalty_term_dynamic_concave_utilities,Laeven+Stadje_2014_Robust_portfolio_choice_indifference_valuation}.
For convex cash-subadditive risk measures, \cite{Mastrogiacomo+RosazzaGianin_2015_Time-consistency_cash-subadditive_risk_measures}
shows that a generalized cocycle condition, formulated along the factorization prescribed by a pair of a discount factor and a probability measure, is sufficient for time-consistency under additional conditions.
In discrete time and under a strictly stronger notion of time-consistency, \cite[Theorem~2]{Mastrogiacomo+RosazzaGianin_2019_Time-consistency_risk_measures_how_strong_is_property} shows that only a one-step version of this condition is necessary.
In this paper, we fully characterize time-consistency of convex cash-subadditive risk measures, in continuous time, in terms of the minimal penalty.

The dual variables of a convex cash-subadditive risk measure are discounted probabilities \cite{ElKaroui+Ravanelli_2009_Cash_sub-additive_risk_measures_interest_rate_ambiguity,Mastrogiacomo+RosazzaGianin_2015_Time-consistency_cash-subadditive_risk_measures}.
We write them as $\F_s$-conditional sub-probability densities, so that, under continuity from above and with $c_{s,t}$ the minimal penalty,
\begin{align}
    \r_{s,t}(X)&=\esssup_{Z\in\mathcal Z_{s,t}}\big\{\E[-ZX|\F_s]-c_{s,t}(Z)\big\},\\
    \mathcal Z_{s,t}&:=\big\{Z\in L^1(\F_t)\, :\,  Z\geq0,\ \E[Z|\F_s]\leq1\big\}.
\end{align}
The set $\mathcal Z_{s,t}$ has the property of being closed under products: $Z^1Z^2\in\mathcal Z_{s,t}$ for $Z^1\in\mathcal Z_{s,u}$ and $Z^2\in\mathcal Z_{u,t}$.
This product structure makes visible the feature that distinguishes the cash-subadditive case.
A probability density factorizes uniquely into its restriction to $\F_u$ and a conditional density given $\F_u$, whereas a sub-probability density $Z$ factorizes as $Z=Z^1Z^2$ in many ways, since its discount can be allocated to the periods $[s,u]$ and $[u,t]$ in many ways.
No allocation is canonical: the discounted expectation $\r_{s,t}(X)=-e^{-r(t-s)}\E[X|\F_s]$, $r>0$, is time-consistent, and its penalty satisfies the cocycle property along the allocation $e^{-r(t-s)}=e^{-r(u-s)}e^{-r(t-u)}$, but not along the restriction factorization, which assigns the whole discount to $[s,u]$.
We therefore replace the cocycle property with the \emph{infimal cocycle property} \itemref{RM:IC}
\[
    c_{s,t}(Z)=\essinf_{Z=Z^1Z^2}\Big\{c_{s,u}(Z^1)+\E\big[Z^1c_{u,t}(Z^2)\big|\F_s\big]\Big\},
\]
where the infimum runs over all factorizations with $Z^1\in\mathcal Z_{s,u}$ and $Z^2\in\mathcal Z_{u,t}$.
In words, the penalty of reaching the density $Z$ at time $t$ is the cheapest way of first reaching an intermediate density $Z^1$ at time $u$ and then the remaining density $Z^2$ at time $t$, the cost of the second period being weighted by $Z^1$.

Our main result, Theorem~\ref{THM:CS-triad}, concerns local, convex, cash-subadditive and normalized risk measures on $L^\infty$.
Under continuity from below, time-consistency and
the infimal cocycle property
are equivalent, and the infimum is attained.
On $L^p$, $p<\infty$, the equivalence holds without any continuity assumption
(Theorem~\ref{THM:CS-triad-p}).
In the cash-additive case, the infimum is attained at the restriction factorization, and
\itemref{RM:IC}
reduces to the classical cocycle property (Remark~\ref{REM:CS-IC-CA}).

The main technical difficulty is the intricate implication from time-consistency to
\itemref{RM:IC}.
In the cash-additive case it follows from the acceptance sets: the factorization is unique, and the decomposition $\A_{s,t}=\A_{s,u}+\A_{u,t}$ splits the supremum defining $c_{s,t}$ into those defining $c_{s,u}$ and $c_{u,t}$.
In the cash-subadditive case the factorization is not unique, and the level decomposition, being a union over levels rather than a Minkowski sum, does not split the penalty.
The proof combines four ingredients.
First, we reduce to densities $Z$ whose penalty has finite expectation, and write $\E[c_{s,t}(Z)]$ as a sup-inf over positions $X$ and intermediate densities $Z^1$.
Second, under continuity from below, the sublevel sets $\{Z^1\, :\,  \E[c_{s,u}(Z^1)]\leq\kappa\}$ of the expected penalty are
weakly compact by the Dunford--Pettis theorem.
Third, on these compact sets, Fan's minimax theorem \cite{Fan_1953_Minimax_theorems} allows us to exchange the supremum and the infimum, for positions in balls of $L^\infty$.
Finally, as the radius of the balls tends to infinity, the Eberlein--\v{S}mulian theorem yields a weak limit $Z^{1}$ of the approximate minimizers, which attains the infimum in \itemref{RM:IC}.
On $L^p$, $p<\infty$, the results follow by restriction to bounded positions, since continuity from below is automatic \cite{Ruszczynski+Shapiro_2006_Optimization_Convex_Risk_Functions,Biagini+Frittelli_2009_Extension_Namioka-Klee_theorem_Fatou_property_risk_measures}.

As an application, we show in Section~\ref{sec:bsde} that, for risk measures induced by BSDEs, the minimal penalty is the value of a stochastic control problem in which an ambiguous interest rate is optimally scheduled over time, and \itemref{RM:IC} is its dynamic programming principle.

The paper is organized as follows.
Section~\ref{sec:prel} fixes the setting.
Section~\ref{sec: tc condition} contains the duality on $L^\infty$ (Section~\ref{SUBSEC:CS-duality}), factorizations and the infimal cocycle property (Section~\ref{SUBSEC:CS-factorization}), the main result (Section~\ref{SUBSEC:CS-main}) and its extension to $L^p$ (Section~\ref{SUBSEC:CS-p}).
Section~\ref{sec:bsde} applies the results to risk measures induced by BSDEs.
Appendices~\ref{APP:CS-p-duality} and~\ref{APP:BSDE-penalty} contain the proofs of Propositions~\ref{PROP:CS-p-duality} and~\ref{PROP:BSDE-penalty}.

\section{Preliminaries}
\label{sec:prel}

Fix a filtered probability space $(\Omega,\F,\mathbb F,\P)$ with filtration $\mathbb F=(\F_t)_{t\geq0}$.
For $t\geq0$ and $p\in[1,+\infty]$, we write $L^p(\F_t):=L^p(\Omega,\F_t,\P)$, and $L^p_+(\F_t)$ and $L^p_-(\F_t)$ for the sets of its nonnegative and nonpositive elements, respectively; $L^0\big(\F_t;[0,+\infty]\big)$ denotes the set of $\F_t$-measurable random variables with values in $[0,+\infty]$.
Random variables that coincide $\P$-a.s.\ are identified, and equalities and inequalities between random variables are understood $\P$-a.s., unless otherwise stated.
Essential infima and suprema are taken with the conventions $\essinf\varnothing=+\infty$ and $\esssup\varnothing=-\infty$; see \cite[Section~A.5]{Follmer+Schied_2016_Stochastic_finance}.
An essential supremum (infimum) that is attained is written $\essmax$ ($\essmin$).

A \emph{(dynamic) risk measure on $L^p$} is a family $\r=(\r_{s,t})_{0\leq s\leq t}$ of monotone maps $\r_{s,t}:L^p(\F_t)\to L^p(\F_s)$, i.e., $X\leq Y$ implies $\r_{s,t}(X)\geq\r_{s,t}(Y)$, for all $X,Y\in L^p(\F_t)$.
The static case is recovered for $s=0$ and trivial $\F_0$, with fixed $t>0$.

Throughout, we shall always assume that $\r$ is
\begin{itemize}[nosep, leftmargin=*, label=$\tinybullet$]
    \item \proplabel{CSA}{RM:CSA}
        \emph{cash-subadditive}, i.e., $\r_{s,t}(X-k)\leq\r_{s,t}(X)+k$ for all $k\in L^p_+(\F_s)$;
    \item \proplabel{N}{RM:N}
        \emph{normalized}, i.e., $\r_{s,t}(0)=0$;
    \item \proplabel{Cx}{RM:Cx}
        \emph{convex}, i.e., $\r_{s,t}\big(\l X+(1-\l)Y\big)\leq\l\r_{s,t}(X)+(1-\l)\r_{s,t}(Y)$ for all $\l\in[0,1]$, $X,Y\in L^p(\F_t)$;
    \item \proplabel{LP}{RM:LP}
        \emph{local}, i.e., $\r_{s,t}(\1_AX+\1_{A^c}Y)=\1_A\r_{s,t}(X)+\1_{A^c}\r_{s,t}(Y)$ for all $A\in\F_s$, $X,Y\in L^p(\F_t)$.
\end{itemize}
For $X\in L^\infty(\F_t)$, let $k:=\|X\|_{L^\infty}\in L^p_+(\F_s)$, so that $-k\leq X\leq k$.
Monotonicity, cash-subadditivity and normalization give
\begin{equation}
\label{EQ:CS-basic}
    |\r_{s,t}(X)|\leq\|X\|_{L^\infty},
    \qquad
    0\leq s\leq t,\ X\in L^\infty(\F_t).
\end{equation}

Moreover, we will say that $\r$ is
\begin{itemize}[nosep, leftmargin=*, label=$\smallbullet$]
    \item \proplabel{TC}{RM:TC}
        \emph{time-consistent} if $\r_{s,t}(X)=\r_{s,u}\bigl(-\r_{u,t}(X)\bigr)$ for all $0\leq s\leq u\leq t$ and $X\in L^p(\F_t)$;
    \item \proplabel{CA}{RM:CA}
        \emph{cash-additive} if $\r_{s,t}(X-k)=\r_{s,t}(X)+k$ for all $k\in L^p(\F_s)$;
    \item \proplabel{C$\downarrow$}{RM:Cabove}
        \emph{continuous from above} if $\r_{s,t}(X_n)\uparrow\r_{s,t}(X)$ $\P$-a.s.\ whenever $X_n\downarrow X$ $\P$-a.s.\ with $X_n,X\in L^p(\F_t)$;
    \item \proplabel{C$\uparrow$}{RM:Cbelow}
        \emph{continuous from below} if $\r_{s,t}(X_n)\downarrow\r_{s,t}(X)$ $\P$-a.s.\ whenever $X_n\uparrow X$ $\P$-a.s.\ with $X_n,X\in L^p(\F_t)$.
\end{itemize}
These notions are standard; see, e.g., \cite{Artzner+Delbaen+Eber+Heath_1999_Coherent_measures_risk} and \cite[Chapter~4]{Follmer+Schied_2016_Stochastic_finance} in the static case, \cite{Delbaen_2006_Structure_m-stable_sets_particular_set_risk_neutral_measures,Detlefsen+Scandolo_2005_Conditional_dynamic_convex_risk_measures,Cheridito+Delbaen+Kupper_2006_Dynamic_monetary_risk_measures_bounded_discrete-time_processes,Bion-Nadal_2008_Dynamic_risk_measures_time_consistency__risk_measures_BMO_martingales} and \cite[Chapter~11]{Follmer+Schied_2016_Stochastic_finance} in the dynamic case, \cite{ElKaroui+Ravanelli_2009_Cash_sub-additive_risk_measures_interest_rate_ambiguity,Mastrogiacomo+RosazzaGianin_2015_Time-consistency_cash-subadditive_risk_measures} for cash-subadditivity, \cite{Laeven+Ferrari+Rosazza+Zullino_2025_measuringfinancialresilienceusing, Ferrari+Laeven+Rosazza+Zullino_2026_financialresilienceevaluation} for an application to financial resilience and \cite{DiNunno+RosazzaGianin_2024_Fully_dynamic_risk_measures_horizon_risk_time-consistency_relations_BSDEs_BSVIEs} for horizon risk.

On $L^\infty$, continuity from below implies continuity from above.
Indeed, by \eqref{EQ:CS-basic} and monotone convergence, $X\mapsto\E[\r_{s,t}(X)]$ is a convex, monotone, cash-subadditive  and normalized functional on $L^\infty(\F_t)$ that is continuous from below; hence it is represented on $\sigma$-additive sub-probabilities by \cite[Propositions~1 and~3]{Mastrogiacomo+RosazzaGianin_2015_Time-consistency_cash-subadditive_risk_measures}, and so it is continuous from above by \cite[Theorem~4]{Mastrogiacomo+RosazzaGianin_2015_Time-consistency_cash-subadditive_risk_measures}.
If $X_n\downarrow X$, then $\r_{s,t}(X_n)\uparrow R\leq\r_{s,t}(X)$ by monotonicity, and $\E[R]=\E[\r_{s,t}(X)]$ by dominated convergence and the continuity from above of $X\mapsto\E[\r_{s,t}(X)]$, so that $R=\r_{s,t}(X)$.

\!Without cash-additivity, convexity is strictly stronger than quasi-convexity \cite{Cerreia-Vioglio+Maccheroni+Marinacci+Montrucchio_2011_Risk_measures_rationality_diversification}.
We nevertheless assume convexity, as in \cite{ElKaroui+Ravanelli_2009_Cash_sub-additive_risk_measures_interest_rate_ambiguity,Mastrogiacomo+RosazzaGianin_2015_Time-consistency_cash-subadditive_risk_measures}, because quasi-convexity is not preserved by the composition $\r_{s,u}(-\r_{u,t}(\cdot))$ that defines time-consistency (see, e.g., \cite[Examples~7 and~8]{Mastrogiacomo+RosazzaGianin_2019_Time-consistency_risk_measures_how_strong_is_property}) and does not yield an additive penalty in general \cite{Frittelli+Maggis_2011_Dual_representation_quasi-convex_conditional_maps}.

\section{Time-consistency and the infimal cocycle property}
\label{sec: tc condition}
In Sections~\ref{SUBSEC:CS-duality}--\ref{SUBSEC:CS-main} we take $p=\infty$; Section~\ref{SUBSEC:CS-p} treats $p\in[1,+\infty)$.

\subsection{Duality on \texorpdfstring{$L^\infty$}{L-infinity}}
\label{SUBSEC:CS-duality}

\begin{definition}
\label{DEF:CS-Z}
    For $0\leq s\leq t$, the set of \emph{$\F_s$-conditional sub-probability densities} on $\F_t$ is
    \[
        \mathcal Z_{s,t}:=\big\{Z\in L^1(\F_t)\, :\,  Z\geq0,\ \E[Z|\F_s]\leq1\big\}.
    \]
\end{definition}

For $s=0$ and trivial $\F_0$, $\mathcal Z_{0,t}$ is the set of densities of the sub-probabilities $\mu\ll\P$, $\mu(\Omega)\leq1$, on which convex cash-subadditive risk measures are represented in \cite[Theorem~4.3~(\textit{b})]{ElKaroui+Ravanelli_2009_Cash_sub-additive_risk_measures_interest_rate_ambiguity}; see also \cite[Remark~2 and Theorem~4]{Mastrogiacomo+RosazzaGianin_2015_Time-consistency_cash-subadditive_risk_measures}.
In the dynamic case, \cite{Mastrogiacomo+RosazzaGianin_2015_Time-consistency_cash-subadditive_risk_measures} uses pairs of a discount factor and a probability measure instead; the conditional density formulation is equivalent (Remark~\ref{REM:CS-DQ}), but, unlike the pairs, it is closed under products, which is what the factorizations of Section~\ref{SUBSEC:CS-factorization} rely on.

\begin{lemma}
\label{LEM:CS-Z}
    Let $0\leq s\leq u\leq t$.
    \begin{enumerate}[nosep, label=$(\alph*)$, leftmargin=*]
        \item $\mathcal Z_{s,t}$ is convex and $\sigma(L^1,L^\infty)$-closed.
        \item $\mathcal Z_{s,t}$ is stable under countable $\F_s$-pasting: if $(A_i)_i\subseteq\F_s$ is a countable partition of $\Omega$ and $Z_i\in\mathcal Z_{s,t}$, then  $\sum_i\1_{A_i}Z_i\in\mathcal Z_{s,t}$.
        \item $\mathcal Z_{s,t}$ is closed under restriction and product:
        \[
            Z\in\mathcal Z_{s,t}\ \Longrightarrow\ \E[Z|\F_u]\in\mathcal Z_{s,u},\qquad
            Z^1\in\mathcal Z_{s,u},\ Z^2\in\mathcal Z_{u,t}\ \Longrightarrow\ Z^1Z^2\in\mathcal Z_{s,t}.
        \]
    \end{enumerate}
\end{lemma}
\begin{proof}
    Convexity and stability under $\F_s$-pasting follow from the properties of conditional expectation.
    If $Z_j\in\mathcal Z_{s,t}$ and $Z_j\to Z$ in $L^1$, then $Z\geq0$ and $\E[Z|\F_s]=\lim_j\E[Z_j|\F_s]\leq1$, by the $L^1$-continuity of the conditional expectation; hence $\mathcal Z_{s,t}$ is closed in $L^1$ and, being convex, $\sigma(L^1,L^\infty)$-closed by Mazur's theorem (see \cite[Theorem~A.63]{Follmer+Schied_2016_Stochastic_finance}).
    The first implication in (\textit{c}) follows from the tower property.
    For the second one, $Z^1Z^2\geq0$ and, by monotone convergence and the tower property, $\E[Z^1Z^2|\F_s]=\E\big[Z^1\E[Z^2|\F_u]\big|\F_s\big]\leq\E[Z^1|\F_s]\leq1$; in particular, $Z^1Z^2\in L^1(\F_t)$.
\end{proof}

\begin{remark}
\label{REM:CS-DQ}
    In \cite[Proposition~6]{Mastrogiacomo+RosazzaGianin_2015_Time-consistency_cash-subadditive_risk_measures} the dual variables are pairs $(D,\Q)$ of a discount factor $D\in L^\infty(\F_s;[0,1])$ and a probability measure $\Q\in\mathcal Q^{\mathrm{ac}}_{s,t}$, the set of probability measures $\Q$ on $(\Omega,\F_t)$ with $\Q\ll\P$ on $\F_t$ and $\Q=\P$ on $\F_s$.
    For $Z\in\mathcal Z_{s,t}$, let $D:=\E[Z|\F_s]$ and let $\Q\in\mathcal Q^{\mathrm{ac}}_{s,t}$ be given by $\!\d\Q/\!\d\P:=(Z/D)\1_{\{D>0\}}+\1_{\{D=0\}}$.
    Then $Z=D\,\!\d\Q/\!\d\P$ and $\E[ZX|\F_s]=D\,\E_\Q[X|\F_s]$ for $X\in L^\infty(\F_t)$; conversely, every such pair defines an element of $\mathcal Z_{s,t}$.
    Working with $Z$ removes the non-uniqueness of the pair on $\{D=0\}$ and makes the products in Lemma~\ref{LEM:CS-Z}~(\textit{c}) available.
\end{remark}

For $Z\in\mathcal Z_{s,t}$ we define the penalty
\begin{equation}
\label{EQ:CS-penalty}
    c_{s,t}(Z)
    :=\esssup_{X\in L^\infty(\F_t)}\big\{\E[-ZX|\F_s]-\r_{s,t}(X)\big\}.
\end{equation}
If $\r_{s,t}$ is cash-additive, then $c_{s,t}(Z)=+\infty$ on $\{\E[Z|\F_s]<1\}$ (Remark~\ref{REM:CS-IC-CA}), and \eqref{EQ:CS-penalty} reduces to the usual minimal penalty on $\mathcal Q^{\mathrm{ac}}_{s,t}$ (Remark~\ref{REM:CS-DQ}).

\begin{proposition}
\label{PROP:CS-duality}
    Let $\r$ be continuous from above.
    Then, for all $0\leq s\leq t$:
    \begin{enumerate}[label=$(\alph*)$, leftmargin=*]
        \item
            $c_{s,t}:\mathcal Z_{s,t}\to L^0(\F_s;[0,+\infty])$ is countably local: if $(A_i)_i\subseteq\F_s$ is a countable partition of $\Omega$ and $Z_i\in\mathcal Z_{s,t}$, then $c_{s,t}(\sum_i\1_{A_i}Z_i)=\sum_i\1_{A_i}c_{s,t}(Z_i)$.
        \item
            For all $X\in L^\infty(\F_t)$,
            \begin{equation}
            \label{EQ:CS-representation}
                \r_{s,t}(X)=\esssup_{Z\in\mathcal Z_{s,t}}\big\{\E[-ZX|\F_s]-c_{s,t}(Z)\big\},
            \end{equation}
            and $c_{s,t}$ is minimal: if \eqref{EQ:CS-representation} holds with some $\a:\mathcal Z_{s,t}\to L^0(\F_s;[0,+\infty])$ in place of $c_{s,t}$, then $c_{s,t}\leq\a$.
        \item
            For every $X\in L^\infty(\F_t)$ there are $Z_n\in\mathcal Z_{s,t}$ such that
            \[
                -\|X\|_{L^\infty}-1\ \leq\ \E[-Z_nX|\F_s]-c_{s,t}(Z_n)\ \uparrow\ \r_{s,t}(X),
            \]
            and for every $Z\in\mathcal Z_{s,t}$ there are $X_n\in L^\infty(\F_t)$ such that
            \[
                0\ \leq\ \E[-ZX_n|\F_s]-\r_{s,t}(X_n)\ \uparrow\ c_{s,t}(Z).
            \]
            In particular,
            \begin{align}
            \label{EQ:CS-expectation_rho}
                \E[\r_{s,t}(X)]
                &=\sup_{Z\in\mathcal Z_{s,t}}\E\bigl[-ZX-c_{s,t}(Z)\bigr],
                \\
            \label{EQ:CS-expectation_c}
                \E[c_{s,t}(Z)]
                &=\sup_{X\in L^\infty(\F_t)}\E\bigl[-ZX-\r_{s,t}(X)\bigr].
            \end{align}
    \end{enumerate}
\end{proposition}
\begin{proof}
    (\textit{a}). This is immediate from \eqref{EQ:CS-penalty};
    choosing $X=0$ gives $c_{s,t}\geq0$.

    (\textit{b}).
    \cite[Proposition~6]{Mastrogiacomo+RosazzaGianin_2015_Time-consistency_cash-subadditive_risk_measures}, applied to the horizon $t$ and translated through Remark~\ref{REM:CS-DQ}, yields \eqref{EQ:CS-representation}, with the minimal penalty \eqref{EQ:CS-penalty}.
    If \eqref{EQ:CS-representation} holds with $\a$, then $\a(Z)\geq\E[-ZX|\F_s]-\r_{s,t}(X)$ for all $X$, hence $\a(Z)\geq c_{s,t}(Z)$.

    (\textit{c}).
    We reason as in \cite[Lemma~1]{Detlefsen+Scandolo_2005_Conditional_dynamic_convex_risk_measures}.
    Fix $X\in L^\infty(\F_t)$ and set $v_X(Z):=\E[-ZX|\F_s]-c_{s,t}(Z)$ for $Z\in\mathcal Z_{s,t}$, so that $\r_{s,t}(X)=\esssup_{Z\in\mathcal Z_{s,t}}v_X(Z)$ by (\textit{b}).
    For $Z,Z'\in\mathcal Z_{s,t}$, let $A:=\{v_X(Z)\geq v_X(Z')\}\in\F_s$ and $Z'':=\1_AZ+\1_{A^c}Z'\in\mathcal Z_{s,t}$ (Lemma~\ref{LEM:CS-Z}~(\textit{b})).
    By (\textit{a}), $v_X(Z'')=\1_Av_X(Z)+\1_{A^c}v_X(Z')=v_X(Z)\vee v_X(Z')$.
    Hence the family $\{v_X(Z)\}_{Z\in\mathcal Z_{s,t}}$ is directed upward, and \cite[Theorem~A.37~(\textit{b})]{Follmer+Schied_2016_Stochastic_finance} yields $Z_n\in\mathcal Z_{s,t}$ with $v_X(Z_n)\uparrow\r_{s,t}(X)$.
    Moreover, $\r_{s,t}(X)\in L^\infty(\F_s)$ by \eqref{EQ:CS-basic}.

    The lower bound, which allows monotone convergence in \eqref{EQ:CS-expectation_rho} and in the proof of Theorem~\ref{THM:CS-triad}, is obtained by pasting.
    Let $N\in\F_s$ be a $\P$-null set outside of which $v_X(Z_n)\uparrow\r_{s,t}(X)\in\R$.
    The sets $B_n:=N\cup\{v_X(Z_n)\geq\r_{s,t}(X)-1\}\in\F_s$ increase to $\Omega$.
    The countable pasting $\hat Z:=\sum_n\1_{B_n\setminus B_{n-1}}Z_n$ (with $B_0:=\varnothing$) satisfies $v_X(\hat Z)\geq\r_{s,t}(X)-1\geq-\|X\|_{L^\infty}-1$ by (\textit{a}).
    Finally, let $A_n:=\{v_X(Z_n)\geq v_X(\hat Z)\}\in\F_s$ and $Z'_n:=\1_{A_n}Z_n+\1_{A_n^c}\hat Z\in\mathcal Z_{s,t}$.
    By (\textit{a}), $v_X(Z'_n)=v_X(Z_n)\vee v_X(\hat Z)$, which is nondecreasing in $n$, bounded from below by $v_X(\hat Z)\geq-\|X\|_{L^\infty}-1$, and converges to $\r_{s,t}(X)\vee v_X(\hat Z)=\r_{s,t}(X)$, since $v_X(\hat Z)\leq\r_{s,t}(X)$ by (\textit{b}).
    Thus $(Z'_n)_n$ satisfies the first claim.

    The second claim is proved in the same way with $w_Z(X):=\E[-ZX|\F_s]-\r_{s,t}(X)$, $X\in L^\infty(\F_t)$, in place of $v_X$: the family $\{w_Z(X)\}_X$ is directed upward by the locality of $\r_{s,t}$, $c_{s,t}(Z)=\esssup_Xw_Z(X)$ by \eqref{EQ:CS-penalty}, and the lower bound $0$ is obtained by pasting with $X=0$, since $w_Z(0)=-\r_{s,t}(0)=0$.
    \eqref{EQ:CS-expectation_rho} and \eqref{EQ:CS-expectation_c} follow by monotone convergence.
\end{proof}

\begin{lemma}
\label{LEM:CS-compact}
    Let $\r$ be continuous from above and fix $0\leq s\leq t$.
    The function $\mathcal Z_{s,t}\ni Z\mapsto\E[c_{s,t}(Z)]\in[0,+\infty]$ is convex and $\sigma(L^1,L^\infty)$-lower semicontinuous.
    If, moreover, $\r_{s,t}$ is continuous from below, then $K^\kappa_{s,t}:=\{Z\in\mathcal Z_{s,t}\, :\,  \E[c_{s,t}(Z)]\leq\kappa\}$ is convex and $\sigma(L^1,L^\infty)$-compact for every $\kappa\geq0$.
\end{lemma}

\begin{proof}
    By \eqref{EQ:CS-expectation_c}, $\E[c_{s,t}(\cdot)]=\sup_{X\in L^\infty(\F_t)}f_X$ on $\mathcal Z_{s,t}$, where $f_X(Z):=\E[-ZX]-\E[\r_{s,t}(X)]$.
    Each $f_X$ is affine, hence $\E[c_{s,t}(\cdot)]$ is convex.
    Moreover, each $f_X$ is $\sigma(L^1,L^\infty)$-continuous;
    then $\E[c_{s,t}(\cdot)]$ is $\sigma(L^1,L^\infty)$-lower semicontinuous since its sublevel sets $\{\E[c_{s,t}(\cdot)]\leq a\}=\bigcap_X\{f_X\leq a\}$, $a\in\R$, are intersections of closed sets.
    This proves the first claim.

    Since $\mathcal Z_{s,t}$ is convex and $\sigma(L^1,L^\infty)$-closed (Lemma~\ref{LEM:CS-Z}~(\textit{a})), $K^\kappa_{s,t}$, a sublevel set of a convex lower semicontinuous function on $\mathcal Z_{s,t}$, is convex and $\sigma(L^1,L^\infty)$-closed.

    Assume now that $\r_{s,t}$ is continuous from below.
    $K^\kappa_{s,t}$ is bounded in $L^1$, since $\E[Z]=\E\big[\E[Z|\F_s]\big]\leq1$ for $Z\in\mathcal Z_{s,t}$.
    Let $(B_n)_n\subseteq\F_t$ with $\1_{B_n}\downarrow0$ a.s.\ and $\l>0$.
    For $Z\in K^\kappa_{s,t}$, \eqref{EQ:CS-expectation_c} with $X=-\l\1_{B_n}$ gives $\kappa\geq\E[c_{s,t}(Z)]\geq\l\,\E[Z\1_{B_n}]-\E[\r_{s,t}(-\l\1_{B_n})]$, i.e., \[
        \sup_{Z\in K^\kappa_{s,t}}
        \E[Z\1_{B_n}]\leq\frac{\kappa+\E[\r_{s,t}(-\l\1_{B_n})]}{\l}.
    \]
    Since $-\l\leq-\l\1_{B_n}\leq0$, monotonicity, cash-subadditivity and normalization give $0=\r_{s,t}(0)\leq\r_{s,t}(-\l\1_{B_n})\leq\r_{s,t}(-\l)\leq\r_{s,t}(0)+\l=\l$.
    Since $-\l\1_{B_n}\uparrow0$ a.s., continuity from below and normalization give $\r_{s,t}(-\l\1_{B_n})\downarrow0$ a.s., hence $\E[\r_{s,t}(-\l\1_{B_n})]\to0$ by dominated convergence.
    Thus $\limsup_n\sup_{Z\in K^\kappa_{s,t}}\E[Z\1_{B_n}]\leq\kappa/\l$ for every $\l>0$, and letting $\l\to\infty$,
    \begin{equation}
    \label{EQ:CS-UI}
        \lim_n\sup_{Z\in K^\kappa_{s,t}}\E[Z\1_{B_n}]=0.
    \end{equation}
    We claim that $K^\kappa_{s,t}$ is uniformly integrable, i.e., since it is bounded in $L^1$, that $\sup\{\E[Z\1_C]\, :\,  Z\in K^\kappa_{s,t},\ C\in\F_t,\ \P(C)\leq\delta\}\to0$ as $\delta\downarrow0$.
    Otherwise, there would be $\eps>0$, $Z_k\in K^\kappa_{s,t}$ and $C_k\in\F_t$ with $\P(C_k)\leq2^{-k}$ and $\E[Z_k\1_{C_k}]\geq\eps$ for all $k\in\N$.
    Since $\sum_k\P(C_k)<\infty$, the Borel--Cantelli lemma (see, e.g., \cite[Section~2.7]{williams_1991_probability}) gives $\P(\limsup_kC_k)=0$, so that $B_n:=\bigcup_{k\geq n}C_k$ satisfies $\1_{B_n}\downarrow\1_{\limsup_kC_k}=0$ a.s.
    However, $\sup_{Z\in K^\kappa_{s,t}}\E[Z\1_{B_n}]\geq\E[Z_n\1_{B_n}]\geq\E[Z_n\1_{C_n}]\geq\eps$ for all $n$, contradicting \eqref{EQ:CS-UI}.
    By the Dunford--Pettis theorem (see \cite[Theorem~A.70]{Follmer+Schied_2016_Stochastic_finance}), a bounded and uniformly integrable subset of $L^1(\F_t)$ is relatively $\sigma(L^1,L^\infty)$-compact.
    Since $K^\kappa_{s,t}$ is $\sigma(L^1,L^\infty)$-closed, it is $\sigma(L^1,L^\infty)$-compact.
\end{proof}

\subsection{Factorizations and the infimal cocycle property}
\label{SUBSEC:CS-factorization}

\begin{definition}
\label{DEF:CS-factorization}
    Let $0\leq s\leq u\leq t$ and $Z\in\mathcal Z_{s,t}$.
    A \emph{factorization of $Z$ at time $u$} is a pair $(Z^1,Z^2)\in\mathcal Z_{s,u}\times\mathcal Z_{u,t}$ with $Z=Z^1Z^2$; we denote by $\Phi_u(Z)$ the set of such pairs.
    We set, with the convention $0\cdot(+\infty)=0$,
    \begin{align}
    \label{EQ:CS-Gamma}
        \Gamma_{s,u,t}(Z^1,Z^2)
        &:=
        c_{s,u}(Z^1)+\E\big[Z^1c_{u,t}(Z^2)\big|\F_s\big],
        \qquad (Z^1,Z^2)\in\mathcal Z_{s,u}\times\mathcal Z_{u,t},\\
        \g_{s,u,t}(Z)
        &:=
        \essinf_{(Z^1,Z^2)\in\Phi_u(Z)}\Gamma_{s,u,t}(Z^1,Z^2), \qquad Z\in\mathcal Z_{s,t}.
    \end{align}
    Moreover, for $Z\in\mathcal Z_{s,t}$ and $Z^1\in\mathcal Z_{s,u}$ we write
    \[
        Z_u:=\E[Z|\F_u],
        \qquad
        Z\oslash Z^1:=(Z/Z^1)\1_{\{Z^1>0\}}+\1_{\{Z^1=0\}}.
    \]
\end{definition}

\begin{lemma}
\label{LEM:CS-factorization}
    Let $0\leq s\leq u\leq t$ and $Z\in\mathcal Z_{s,t}$.
    \begin{enumerate}[label=$(\alph*)$, leftmargin=*]
        \item
            If $(Z^1,Z^2)\in\Phi_u(Z)$, then $Z^1\geq Z_u$.
        \item
            If $Z^1\in\mathcal Z_{s,u}$ and $Z^1\geq Z_u$, then $(Z^1,Z\oslash Z^1)\in\Phi_u(Z)$.
            Moreover, 
            \[
                \Gamma_{s,u,t}(Z^1,Z^2)=\Gamma_{s,u,t}(Z^1,Z\oslash Z^1), \qquad  (Z^1,Z^2)\in\Phi_u(Z).
            \]
            In particular, $\Phi_u(Z)\neq\varnothing$.
        \item
            If $(A_i)_i\subseteq\F_s$ is a countable partition of $\Omega$, $Z_i\in\mathcal Z_{s,t}$ and $(Z^1_i,Z^2_i)\in\Phi_u(Z_i)$, then $\sum_i(Z^1_i,Z^2_i)\1_{A_i}\in\Phi_u\big(\sum_iZ_i\1_{A_i}\big)$ and $\Gamma_{s,u,t}$ is local with respect to such pastings:
            \[
                \Gamma_{s,u,t}\Big(\sum_i\1_{A_i}(Z^1_i,Z^2_i)\Big)=\sum_i\1_{A_i}\Gamma_{s,u,t}(Z^1_i,Z^2_i).
            \]
    \end{enumerate}
\end{lemma}
\begin{proof}
    (\textit{a}). We have $Z_u=\E[Z^1Z^2|\F_u]=Z^1\E[Z^2|\F_u]\leq Z^1$.

    (\textit{b}).
    Since $\E[Z\1_{\{Z_u=0\}}]=\E[Z_u\1_{\{Z_u=0\}}]=0$ and $Z\geq0$, we have $Z=0$ a.s.\ on $\{Z_u=0\}\supseteq\{Z^1=0\}$, whence $Z=Z^1\cdot(Z\oslash Z^1)$.
    Moreover, $\E[Z\oslash Z^1|\F_u]=Z_u/Z^1\leq1$ on $\{Z^1>0\}$ and $=1$ on $\{Z^1=0\}$, so that $Z\oslash Z^1\in\mathcal Z_{u,t}$.
    If $(Z^1,Z^2)\in\Phi_u(Z)$, then $Z^2=Z\oslash Z^1$ on $\{Z^1>0\}\in\F_u$; by locality of $c_{u,t}$ (immediate from \eqref{EQ:CS-penalty}) and the convention $0\cdot(+\infty)=0$, $Z^1c_{u,t}(Z^2)=Z^1c_{u,t}(Z\oslash Z^1)$.

    (\textit{c}). This follows from Lemma~\ref{LEM:CS-Z}~(\textit{b}), applied to $\mathcal Z_{s,u}$ and $\mathcal Z_{u,t}$ (note $\F_s\subseteq\F_u$), and from the locality of $c_{s,u}$ and $c_{u,t}$.
\end{proof}

By Lemma~\ref{LEM:CS-factorization}, the only free variable in a factorization is the intermediate density $Z^1\geq Z_u$; on $\{Z_u>0\}$ it is determined by the $\F_u$-measurable factor $\E[Z^2|\F_u]=Z_u/Z^1\in(0,1]$, which is the part of the discount allocated to $[u,t]$.

\begin{definition}
\label{DEF:IC}
    The penalty $c$ satisfies the \emph{infimal cocycle property} if $c_{s,t}=\g_{s,u,t}$ for all $0\leq s\leq u\leq t$, namely
    \begin{itemize}[leftmargin=3.6em, labelwidth=3em, labelsep=0.6em, align=left]
        \item[\textup{(IC)}] \proplabel{IC}{RM:IC}
            $\ds c_{s,t}(Z)=\essinf_{(Z^1,Z^2)\in\Phi_u(Z)}\Big\{c_{s,u}(Z^1)+\E\big[Z^1c_{u,t}(Z^2)\big|\F_s\big]\Big\}$, $\qquad Z\in\mathcal Z_{s,t}$.
    \end{itemize}
\end{definition}

\begin{remark}[The restriction factorization]
\label{REM:CS-IC-CA}
    Every $Z\in\mathcal Z_{s,t}$ admits the restriction factorization $(Z_u,Z\oslash Z_u)\in\Phi_u(Z)$, which allocates the whole discount to $[s,u]$.
    If $\r_{u,t}$ is cash-additive, this is the only relevant one, since discounting on $[u,t]$ has infinite cost: $\r_{u,t}(k)=-k$ and \eqref{EQ:CS-penalty} give $c_{u,t}(Z^2)\geq k(1-\E[Z^2|\F_u])$ for all $k\in L^\infty(\F_u)$, hence $c_{u,t}(Z^2)=+\infty$ on $\{\E[Z^2|\F_u]<1\}$, which, by Lemma~\ref{LEM:CS-factorization}, contains the set $\{Z^1>Z_u\}$ where a factorization differs from the restriction one.
    Hence \itemref{RM:IC} reduces to $c_{s,t}(Z)=c_{s,u}(Z_u)+\E[Z_uc_{u,t}(Z\oslash Z_u)|\F_s]$, which, for $Z=\!\d\Q/\!\d\P$ with $\Q\in\mathcal Q^{\mathrm{ac}}_{s,t}$ (Remark~\ref{REM:CS-DQ}), is the classical cocycle property $c_{s,t}(\Q)=c_{s,u}(\Q)+\E_\Q[c_{u,t}(\Q)|\F_s]$; see \cite{Bion-Nadal_2008_Dynamic_risk_measures_time_consistency__risk_measures_BMO_martingales,Bion-Nadal_2009_Time_consistent_dynamic_risk_processes}.

    In the cash-subadditive case the restriction is in general not optimal.
    For instance, for $r>0$ the \emph{discounted expectation} $\r_{s,t}(X):=-e^{-r(t-s)}\E[X|\F_s]$ is a local, convex, cash-subadditive, normalized and time-consistent risk measure on $L^\infty$, continuous from above and from below, but not cash-additive.
    Since $\r_{s,t}$ is linear, \eqref{EQ:CS-penalty} gives $c_{s,t}(e^{-r(t-s)})=0$ and $c_{s,t}(Z)=+\infty$ on $\{\P(Z\neq e^{-r(t-s)}|\F_s)>0\}$: only the deterministic discount is admissible.
    For $Z:=e^{-r(t-s)}$ and $s<u<t$, the restriction puts the whole discount on $[s,u]$ and gives $\Gamma_{s,u,t}(Z_u,Z\oslash Z_u)=+\infty$, whereas splitting it as $(e^{-r(u-s)},e^{-r(t-u)})$ gives $\Gamma_{s,u,t}=0=c_{s,t}(Z)$.
    Thus the infimum in \itemref{RM:IC} is genuinely an optimization over factorizations.
\end{remark}

\subsection{Main result on \texorpdfstring{$L^\infty$}{L-infinity}}
\label{SUBSEC:CS-main}

\begin{theorem}
\label{THM:CS-triad}
    Let $\r$ be a local, convex, cash-subadditive and normalized risk measure on $L^\infty$.
    \begin{enumerate}[label=$(\alph*)$, leftmargin=*]
        \item
            If $\r$ is time-consistent, then $c$ satisfies the sub-cocycle inequality: for $0\leq s\leq u\leq t$
            \begin{equation}
            \label{EQ:CS-subcocycle}
                c_{s,t}(Z^1Z^2)\leq c_{s,u}(Z^1)+\E\big[Z^1c_{u,t}(Z^2)\big|\F_s\big],
                \qquad Z^1\in\mathcal Z_{s,u},\ Z^2\in\mathcal Z_{u,t}.
            \end{equation}
        \item
            If $\r$ is continuous from above and $c$ satisfies \itemref{RM:IC}, then $\r$ is time-consistent.
        \item
            If $\r$ is continuous from below, then $\r$ is time-consistent if and only if $c$ satisfies \itemref{RM:IC}.
            In this case,
            the essential infimum in \itemref{RM:IC} is attained.
    \end{enumerate}
\end{theorem}

\begin{proof}
\emph{\textbf{\textup{(\textit{a})}.}}
    Let $0\leq s\leq u\leq t$, $Z^1\in\mathcal Z_{s,u}$, $Z^2\in\mathcal Z_{u,t}$ and $X\in L^\infty(\F_t)$.
    By time-consistency and \eqref{EQ:CS-penalty} for $c_{s,u}$ at $-\r_{u,t}(X)\in L^\infty(\F_u)$,
    \[
        \r_{s,t}(X)=\r_{s,u}\bigl(-\r_{u,t}(X)\bigr)\geq\E\big[Z^1\r_{u,t}(X)\big|\F_s\big]-c_{s,u}(Z^1).
    \]
    Since  $\E[-Z^1Z^2X|\F_s]=\E\big[Z^1\E[-Z^2X|\F_u]\big|\F_s\big]$, adding this identity to the previous inequality yields
    \begin{align}
        \E[-Z^1Z^2X|\F_s]-\r_{s,t}(X)
        &\leq c_{s,u}(Z^1)+\E\Big[Z^1\big(\E[-Z^2X|\F_u]-\r_{u,t}(X)\big)\Big|\F_s\Big]\\
        &\leq c_{s,u}(Z^1)+\E\big[Z^1c_{u,t}(Z^2)\big|\F_s\big],
    \end{align}
    where the second inequality follows from $Z^1\geq0$ and
    \eqref{EQ:CS-penalty} for $c_{u,t}$ (with the convention $0\cdot(+\infty)=0$).
    Taking the essential supremum over $X$ gives \eqref{EQ:CS-subcocycle}.
    \smallskip

\emph{\textbf{\textup{(\textit{b})}.}}
    Assume that $\r$ is continuous from above.
    We first show that, for all $0\leq s\leq u\leq t$ and $X\in L^\infty(\F_t)$,
    \begin{equation}
    \label{EQ:CS-composition}
        \r_{s,u}\bigl(-\r_{u,t}(X)\bigr)=\esssup_{Z\in\mathcal Z_{s,t}}\big\{\E[-ZX|\F_s]-\g_{s,u,t}(Z)\big\}.
    \end{equation}
    Fix $Z^1\in\mathcal Z_{s,u}$.
    By Proposition~\ref{PROP:CS-duality}~(\textit{c}) at $(u,t)$, there are $Z^2_n\in\mathcal Z_{u,t}$ such that $Y_n:=\E[-Z^2_nX|\F_u]-c_{u,t}(Z^2_n)$ increases to $\r_{u,t}(X)$ and $Y_n\geq-\|X\|_{L^\infty}-1$.
    By monotone convergence, since $Z^1\geq0$ and $\E[Z^1\E[-Z^2X|\F_u]|\F_s]=\E[-Z^1Z^2X|\F_s]$,
    \[
        \E\big[Z^1\r_{u,t}(X)\big|\F_s\big]
        =\esssup_{Z^2\in\mathcal Z_{u,t}}\Big\{\E[-Z^1Z^2X|\F_s]-\E\big[Z^1c_{u,t}(Z^2)\big|\F_s\big]\Big\},
    \]
    where ``$\geq$'' holds because $\E[-Z^2X|\F_u]-c_{u,t}(Z^2)\leq\r_{u,t}(X)$ for each $Z^2$.
    Plugging this into \eqref{EQ:CS-representation} for $\r_{s,u}$ at $-\r_{u,t}(X)\in L^\infty(\F_u)$ gives
    \[
        \r_{s,u}\bigl(-\r_{u,t}(X)\bigr)
        =\esssup_{(Z^1,Z^2)\in\mathcal Z_{s,u}\times\mathcal Z_{u,t}}\Big\{\E[-Z^1Z^2X|\F_s]-\Gamma_{s,u,t}(Z^1,Z^2)\Big\}.
    \]
    By Lemmas~\ref{LEM:CS-Z}~(\textit{c}) and~\ref{LEM:CS-factorization}~(\textit{b}), the products $Z^1Z^2$ range over all of $\mathcal Z_{s,t}$ and, for fixed $Z$, the pairs with $Z^1Z^2=Z$ form $\Phi_u(Z)\neq\varnothing$.
    Grouping the pairs according to $Z=Z^1Z^2$ gives \eqref{EQ:CS-composition}.

    Under \itemref{RM:IC}, $\g_{s,u,t}=c_{s,t}$, so \eqref{EQ:CS-composition} and \eqref{EQ:CS-representation} for $\r_{s,t}$ give $\r_{s,u}(-\r_{u,t}(X))=\r_{s,t}(X)$, i.e., $\r$ is time-consistent.
    \smallskip

\emph{\textbf{\textup{(\textit{c})}.}}
    Since continuity from below implies continuity from above (Section~\ref{sec:prel}), (\textit{b}) applies; hence it suffices to show that time-consistency implies \itemref{RM:IC} and that the essential infimum is attained.
    Assume that $\r$ is time-consistent and continuous from below, fix $0\leq s\leq u\leq t$ and $Z\in\mathcal Z_{s,t}$.
    By (\textit{a}), $c_{s,t}(Z)\leq\g_{s,u,t}(Z)$, so it suffices to find $(Z^{1*},Z^{2*})\in\Phi_u(Z)$ with
    \begin{equation}
    \label{EQ:CS-goal}
        \Gamma_{s,u,t}(Z^{1*},Z^{2*})\leq c_{s,t}(Z)\quad\P\text{-a.s.}
    \end{equation}

    \emph{Step 1: reduction to $C:=\E[c_{s,t}(Z)]<\infty$.}
    It suffices to prove \eqref{EQ:CS-goal} for every $Z\in\mathcal Z_{s,t}$ with $\E[c_{s,t}(Z)]<\infty$.
    Indeed, let $Z\in\mathcal Z_{s,t}$ be arbitrary.
    By \eqref{EQ:CS-expectation_rho} with $X=0$, there is $Z_0\in\mathcal Z_{s,t}$ with $\E[c_{s,t}(Z_0)]\leq1$.
    For $k\in\N$ let $A_k:=\{c_{s,t}(Z)\leq k\}\in\F_s$, $A_0:=\varnothing$, and $Z_k:=\1_{A_k}Z+\1_{A_k^c}Z_0$, so that $\E[c_{s,t}(Z_k)]\leq k+1$ by locality and $c_{s,t}(Z_k)=c_{s,t}(Z)$ on $A_k$.
    By the reduced claim applied to $Z_k$, there is $(Z^{1*}_k,Z^{2*}_k)\in\Phi_u(Z_k)$ with $\Gamma_{s,u,t}(Z^{1*}_k,Z^{2*}_k)\leq c_{s,t}(Z_k)$.
    By Lemma~\ref{LEM:CS-factorization}~(\textit{c}),
    \[
        (Z^{1*},Z^{2*}):=\sum_{k=1}^\infty(Z^{1*}_k,Z^{2*}_k)\1_{A_k\setminus A_{k-1}}+(Z_u,Z\oslash Z_u)\1_{\{c_{s,t}(Z)=+\infty\}}\in\Phi_u(Z)
    \]
    and satisfies \eqref{EQ:CS-goal}.
    \smallskip

\emph{Step 2: a saddle function.}
    For $X\in L^\infty(\F_t)$ and $Z^1\in\mathcal Z_{s,u}$ let
    \[
        f(X,Z^1):=\E[-ZX]-\E\big[Z^1\r_{u,t}(X)\big]+\E[c_{s,u}(Z^1)]\in(-\infty,+\infty].
    \]
    By \eqref{EQ:CS-expectation_c} for $c_{s,t}$, by time-consistency, and by \eqref{EQ:CS-expectation_rho} for $\r_{s,u}$ at $-\r_{u,t}(X)$,
    \begin{equation}
    \label{EQ:CS-supinf}
        C=\sup_{X\in L^\infty(\F_t)}\Big\{\E[-ZX]-\E\big[\r_{s,u}\bigl(-\r_{u,t}(X)\bigr)\big]\Big\}
        =\sup_{X\in L^\infty(\F_t)}\ \inf_{Z^1\in\mathcal Z_{s,u}}f(X,Z^1).
    \end{equation}

\emph{Step 3: the inner supremum.}
    We claim that, for $Z^1\in\mathcal Z_{s,u}$,
    \begin{equation}
    \label{EQ:CS-innersup}
        \sup_{X\in L^\infty(\F_t)}f(X,Z^1)=
        \begin{cases}
            \E\big[\Gamma_{s,u,t}(Z^1,Z\oslash Z^1)\big], & \text{if }Z^1\geq Z_u,\\
            +\infty, & \text{otherwise}.
        \end{cases}
    \end{equation}
    If $Z^1\geq Z_u$, set $Z^2:=Z\oslash Z^1$, so that $Z=Z^1Z^2$ and
    \[
        f(X,Z^1)=\E\Big[Z^1\big(\E[-Z^2X|\F_u]-\r_{u,t}(X)\big)\Big]+\E[c_{s,u}(Z^1)].
    \]
    By Proposition~\ref{PROP:CS-duality}~(\textit{c}) at $(u,t)$, there are $X_n\in L^\infty(\F_t)$ such that the bracket is nonnegative and increases to $c_{u,t}(Z^2)$, its upper bound by \eqref{EQ:CS-penalty}; monotone convergence gives the first case.
    Otherwise, $B:=\{Z^1<Z_u\}\in\F_u$ has positive probability.
    For $\l>0$, locality, normalization and cash-subadditivity give $\r_{u,t}(-\l\1_B)=\1_B\r_{u,t}(-\l)\leq\l\1_B$; since $B\in\F_u$, $\E[Z\1_B]=\E[Z_u\1_B]$, so that
    \[
        f(-\l\1_B,Z^1)\geq\l\,\E\big[(Z_u-Z^1)\1_B\big]+\E[c_{s,u}(Z^1)]\xrightarrow[\l\to\infty]{}+\infty.
    \]
    \smallskip

    \emph{Step 4: minimax on balls.}
    For $r>0$ let $\mathbb B_r:=\{X\in L^\infty(\F_t)\, :\,  \|X\|_{L^\infty}\leq r\}$.
    For $X\in\mathbb B_r$, $|\r_{u,t}(X)|\leq r$ by \eqref{EQ:CS-basic} and $\E[Z^1]\leq1$ give $f(X,Z^1)\geq\E[-ZX]-r+\E[c_{s,u}(Z^1)]$.
    By \eqref{EQ:CS-expectation_rho} with $X=0$ there is $Z^1_0\in\mathcal Z_{s,u}$ with $\E[c_{s,u}(Z^1_0)]\leq1$, and $f(X,Z^1_0)\leq\E[-ZX]+r+1$.
    Hence $f(X,Z^1)\le f(X,Z^1_0)$ implies $\E[c_{s,u}(Z^1)]-r\leq r+1$, i.e., $Z^1\in K^{2r+1}_{s,u}$.
    Since also $Z^1_0\in K^{2r+1}_{s,u}$, every $Z^1\notin K^{2r+1}_{s,u}$ satisfies $f(X,Z^1)>f(X,Z^1_0)\geq\inf_{K^{2r+1}_{s,u}}f(X,\cdot)$, so that
    \[
        \inf_{Z^1\in\mathcal Z_{s,u}}f(X,Z^1)=\inf_{Z^1\in K^{2r+1}_{s,u}}f(X,Z^1),\qquad X\in\mathbb B_r.
    \]
    On $K^{2r+1}_{s,u}\times\mathbb B_r$, $f$ is real-valued and concave in $X$, since $\r_{u,t}$ is convex and $Z^1\geq0$.
    It is convex and $\sigma(L^1,L^\infty)$-lower semicontinuous in $Z^1$, by Lemma~\ref{LEM:CS-compact} and since $\r_{u,t}(X)\in L^\infty(\F_u)$.
    Since $K^{2r+1}_{s,u}$ is compact by Lemma~\ref{LEM:CS-compact}, Fan's minimax theorem \cite[Theorem~2]{Fan_1953_Minimax_theorems} yields, with $H_r(Z^1):=\sup_{X\in\mathbb B_r}f(X,Z^1)$,
    \[
        \sup_{X\in\mathbb B_r}\inf_{Z^1\in\mathcal Z_{s,u}}f(X,Z^1)
        =\min_{Z^1\in K^{2r+1}_{s,u}}H_r(Z^1)
        \geq h(r):=\inf_{Z^1\in\mathcal Z_{s,u}}H_r(Z^1).
    \]
    By \eqref{EQ:CS-supinf}, the left-hand side is at most $C$; hence $h(r)\leq C$ for all $r>0$.
    \smallskip

    \emph{Step 5: letting $r\to\infty$.}
    Since $\r_{u,t}(0)=0$, $H_r(Z^1)\geq f(0,Z^1)=\E[c_{s,u}(Z^1)]$, so that $\{H_r\leq C+1\}\subseteq K^{C+1}_{s,u}$.
    $H_r$ is lower semicontinuous (a supremum of such maps) and $K^{C+1}_{s,u}$ is compact, so $h(r)$ is attained at some $Z^1_r\in K^{C+1}_{s,u}$.
    By the Eberlein--\v{S}mulian theorem (see \cite[Theorem~A.69]{Follmer+Schied_2016_Stochastic_finance}) there are $r_j\uparrow\infty$ and $Z^{1*}\in K^{C+1}_{s,u}$ with $Z^1_{r_j}\to Z^{1*}$ in $\sigma(L^1,L^\infty)$.
    Since $H_r$ is nondecreasing in $r$, for fixed $r$ and $r_j\geq r$ we have $H_r(Z^1_{r_j})\leq H_{r_j}(Z^1_{r_j})=h(r_j)\leq C$, and lower semicontinuity gives $H_r(Z^{1*})\leq C$.
    Taking the supremum over $r$, since $\bigcup_r\mathbb B_r=L^\infty(\F_t)$, and using \eqref{EQ:CS-innersup}, we get $Z^{1*}\geq Z_u$ and, with $Z^{2*}:=Z\oslash Z^{1*}$,
    \[
        \E\big[\Gamma_{s,u,t}(Z^{1*},Z^{2*})\big]=\sup_{X\in L^\infty(\F_t)}f(X,Z^{1*})\leq C=\E[c_{s,t}(Z)].
    \]

    \emph{Step 6: conclusion.}
    By Lemma~\ref{LEM:CS-factorization}~(\textit{b}), $(Z^{1*},Z^{2*})\in\Phi_u(Z)$, and by \eqref{EQ:CS-subcocycle}, $c_{s,t}(Z)\leq\Gamma_{s,u,t}(Z^{1*},Z^{2*})$ a.s.
    By Step~5 and $C<\infty$, $\E[\Gamma_{s,u,t}(Z^{1*},Z^{2*})-c_{s,t}(Z)]\leq0$, hence $\Gamma_{s,u,t}(Z^{1*},Z^{2*})=c_{s,t}(Z)$ a.s.
    In particular, \eqref{EQ:CS-goal} holds (with equality).
\end{proof}

\begin{remark}[Relation with \cite{Mastrogiacomo+RosazzaGianin_2015_Time-consistency_cash-subadditive_risk_measures,Mastrogiacomo+RosazzaGianin_2019_Time-consistency_risk_measures_how_strong_is_property}]
\label{REM:CS-MRG}
    A family of pairs $(D_{s,t},\Q)$ as in Remark~\ref{REM:CS-DQ} prescribes a factorization.
    Indeed, let $L_v:=\E[\!\d\Q/\!\d\P|\F_v]$ and $L_{v,w}:=L_w/L_v$; by Bayes' formula, the stability assumption (QDa) of \cite{Mastrogiacomo+RosazzaGianin_2019_Time-consistency_risk_measures_how_strong_is_property} with $\Q_1=\Q_2=\Q$ and $D^1=D^2=D$ requires $Z:=D_{s,t}L_{s,t}$ to factorize as $Z=Z^1Z^2$, with $Z^1:=D_{s,u}L_{s,u}$ and $Z^2:=D_{u,t}L_{u,t}$.
    Along this factorization, their generalized cocycle condition (Cc) for the minimal penalty reads $c_{s,t}(Z)=\Gamma_{s,u,t}(Z^1,Z^2)$, and \cite[Theorem~1]{Mastrogiacomo+RosazzaGianin_2019_Time-consistency_risk_measures_how_strong_is_property} shows that, together with the other assumptions of that theorem, it implies time-consistency.
    In discrete time and under continuity from below, \cite[Theorem~2]{Mastrogiacomo+RosazzaGianin_2019_Time-consistency_risk_measures_how_strong_is_property} shows that a one-step version of (Cc), with $u=s+1$ and horizon $t=T$, holds along a fixed factorization if $\big(\r_{s,t}(\cdot+X_s)+X_s\big)_{s\leq t}$ is time-consistent for every adapted $(X_s)_{0\leq s\leq t}$, a property that is strictly stronger than time-consistency.
    Theorem~\ref{THM:CS-triad} provides the converse under time-consistency alone, with the prescribed factorization replaced by an optimal one.
    By Remark~\ref{REM:CS-IC-CA}, (Cc) along a fixed factorization, such as the restriction one, may fail for a time-consistent $\r$; e.g., the discounted expectation is time-consistent but does not satisfy the stronger property above.
\end{remark}

\subsection{The case \texorpdfstring{$p<\infty$}{p finite}}
\label{SUBSEC:CS-p}

Throughout this subsection, let $p\in[1,+\infty)$.
Unlike in Sections~\ref{SUBSEC:CS-duality}--\ref{SUBSEC:CS-main}, no continuity is assumed: the automatic continuity of finite convex monotone functionals on Banach lattices (see \cite{Ruszczynski+Shapiro_2006_Optimization_Convex_Risk_Functions,Biagini+Frittelli_2009_Extension_Namioka-Klee_theorem_Fatou_property_risk_measures}) yields the Lebesgue property of $\r$, hence continuity from above and from below, for free (Lemma~\ref{LEM:CS-p-basic}~(\textit{a})).
As a consequence, the restriction of $\r$ to bounded positions satisfies the assumptions of Sections~\ref{SUBSEC:CS-duality}--\ref{SUBSEC:CS-main}.
This is how the results on $L^\infty$
are transferred to $L^p$.

\begin{lemma}
\label{LEM:CS-p-basic}
    For all $0\leq s\leq t$:
    \begin{enumerate}[label=$(\alph*)$, leftmargin=*]
        \item
            \begin{itemize}[nosep, leftmargin=*, label=\scriptsize$\bullet$]
                \item $L^p(\F_t)\ni X\mapsto\E[\1_A\r_{s,t}(X)]$ is norm-continuous for every $A\in\F_s$;
                \item $\r$ has the Lebesgue property: $X_n\to X$ $\P$-a.s.\ and $|X_n|\leq U\in L^p(\F_t)$ imply $\r_{s,t}(X_n)\to\r_{s,t}(X)$ $\P$-a.s.
            \end{itemize}
        \item
            The restriction $\tilde\r$ of $\r$ to $L^\infty$
            is a local, convex, cash-subadditive and normalized risk measure on $L^\infty$, continuous from above and from below.
    \end{enumerate}
\end{lemma}
\begin{proof}
    Fix $0\leq s\leq t$.

    (\textit{a}).
    \emph{Norm continuity.}
    Let $A\in\F_s$ and $\phi_A(X):=\E[\1_A\r_{s,t}(X)]$, $X\in L^p(\F_t)$.
    Since $\r_{s,t}(X)\in L^p(\F_s)\subseteq L^1(\F_s)$, $\phi_A$ is real-valued; since $\r_{s,t}$ is convex and monotone and $\1_A\geq0$, $\phi_A$ is convex and nonincreasing.
    Hence $\phi_A$ is norm-continuous on the Banach lattice $L^p(\F_t)$ by \cite[Proposition~3.1]{Ruszczynski+Shapiro_2006_Optimization_Convex_Risk_Functions}, an extension of the Namioka--Klee theorem; see also \cite{Biagini+Frittelli_2009_Extension_Namioka-Klee_theorem_Fatou_property_risk_measures}.

    \emph{Lebesgue property.}
    Let first $X_n,X\in L^p(\F_t)$ with $X_n\downarrow X$ $\P$-a.s.\ (the case $X_n\uparrow X$ is symmetric).
    Since $|X_n-X|\leq|X_1-X|\in L^p$ and $p<\infty$, $X_n\to X$ in $L^p$ by dominated convergence.
    By monotonicity, $\r_{s,t}(X_n)$ is nondecreasing and bounded above by $\r_{s,t}(X)$, so it converges $\P$-a.s.\ to some $R$ with $\r_{s,t}(X_1)\leq R\leq\r_{s,t}(X)$; in particular, $R$ is integrable.
    By monotone convergence and the continuity of $\phi_\Omega$, $\E[R]=\lim_n\E[\r_{s,t}(X_n)]=\E[\r_{s,t}(X)]$.
    Since $\r_{s,t}(X)-R\geq0$ has zero expectation, $R=\r_{s,t}(X)$ $\P$-a.s.

    Let now $X_n\to X$ $\P$-a.s.\ with $|X_n|\leq U\in L^p(\F_t)$, and set $I_n:=\inf_{k\geq n}X_k$ and $S_n:=\sup_{k\geq n}X_k$.
    Then $|I_n|,|S_n|,|X|\leq U$, so that $I_n,S_n,X\in L^p(\F_t)$, and $I_n\uparrow X$, $S_n\downarrow X$ $\P$-a.s.
    By monotonicity, $\r_{s,t}(S_n)\leq\r_{s,t}(X_n)\leq\r_{s,t}(I_n)$, and both bounds converge to $\r_{s,t}(X)$ $\P$-a.s.\ by the monotone case.
    Hence $\r_{s,t}(X_n)\to\r_{s,t}(X)$ $\P$-a.s.
    Continuity from above and from below are the special cases $X_n\downarrow X$ and $X_n\uparrow X$, with $U:=|X_1|+|X|$.
    \smallskip

    (\textit{b}).
    By \eqref{EQ:CS-basic}, $\r_{s,t}(X)\in L^\infty(\F_s)$ for $X\in L^\infty(\F_t)$, so that $\tilde\r$ is a dynamic risk measure on $L^\infty$.
    Locality, convexity and normalization pass to the restriction, and so does cash-subadditivity, since $L^\infty_+(\F_s)\subseteq L^p_+(\F_s)$.
    If $X_n,X\in L^\infty(\F_t)$ and $X_n\downarrow X$ or $X_n\uparrow X$ $\P$-a.s., then $|X_n|\leq|X_1|+|X|\in L^p(\F_t)$, so that the Lebesgue property in (\textit{a}) gives continuity from above and from below of $\tilde\r$.
\end{proof}

We keep the dual variables $\mathcal Z_{s,t}\subseteq L^1(\F_t)$ of Section~\ref{SUBSEC:CS-duality}, and we define $c_{s,t}(Z)$ by \eqref{EQ:CS-penalty}, i.e., through bounded positions only.
By Lemma~\ref{LEM:CS-p-basic}~(\textit{b}), $c$ is the minimal penalty of the restriction $\tilde\r$ of $\r$ to $L^\infty$, so that all the results of Sections~\ref{SUBSEC:CS-duality}--\ref{SUBSEC:CS-main} apply to $\tilde\r$ and $c$.
The next result shows that $c$ is also the minimal penalty of $\r$ on the whole of $L^p$, so that the penalty does not depend on $p$.

\begin{proposition}
\label{PROP:CS-p-duality}
    Let $0\leq s\leq t$.
    For all $Z\in\mathcal Z_{s,t}$ and $X\in L^p(\F_t)$, we have $\E[Z|X||\F_s]<\infty$ on $\{c_{s,t}(Z)<\infty\}$, so that $\E[-ZX|\F_s]-c_{s,t}(Z)$ is well defined, with the convention that it equals $-\infty$ on $\{c_{s,t}(Z)=+\infty\}$.
    Moreover, the dual representation \eqref{EQ:CS-representation} holds for all $X\in L^p(\F_t)$, and $c_{s,t}$ is minimal.
\end{proposition}

Since Proposition~\ref{PROP:CS-p-duality} is not used in what follows, we defer its proof to Appendix~\ref{APP:CS-p-duality}.

\begin{theorem}
\label{THM:CS-triad-p}
    Let $p\in[1,+\infty)$ and let $\r$ be a local, convex, cash-subadditive and normalized risk measure on $L^p$.
    \begin{enumerate}[label=$(\alph*)$, leftmargin=*]
        \item
            If $\r$ is time-consistent, then $c$ satisfies the sub-cocycle inequality \eqref{EQ:CS-subcocycle}.
        \item
            $\r$ is time-consistent if and only if $c$ satisfies \itemref{RM:IC}.
            In this case, the essential infimum in \itemref{RM:IC} is attained.
    \end{enumerate}
\end{theorem}
\begin{proof}
    Let $\tilde\r$ be the restriction of $\r$ to $L^\infty$, as in Lemma~\ref{LEM:CS-p-basic}~(\textit{b}).
    By the latter, $\tilde\r$ is a local, convex, cash-subadditive and normalized risk measure on $L^\infty$, continuous from above and from below, so that all parts of Theorem~\ref{THM:CS-triad} apply to $\tilde\r$.
    Moreover, by \eqref{EQ:CS-penalty}, the penalty of $\tilde\r$ is $c$.

    We first show that $\r$ is time-consistent if and only if $\tilde\r$ is time-consistent.
    The forward implication holds because
    $\r_{u,t}(X)\in L^\infty(\F_u)$ for $X\in L^\infty(\F_t)$ by \eqref{EQ:CS-basic}.
    Conversely, assume that $\tilde\r$ is time-consistent, let $0\leq s\leq u\leq t$ and $X\in L^p(\F_t)$, and set $X_k:=(X\wedge k)\vee(-k)$, $k\in\N$.
    Then $X_k\in L^\infty(\F_t)$, so that $\r_{s,t}(X_k)=\r_{s,u}(-\r_{u,t}(X_k))$ for every $k$.
    Since $X_k\to X$ $\P$-a.s.\ and $|X_k|\leq|X|$, the Lebesgue property in Lemma~\ref{LEM:CS-p-basic}~(\textit{a}) gives $\r_{s,t}(X_k)\to\r_{s,t}(X)$ and $\r_{u,t}(X_k)\to\r_{u,t}(X)$ $\P$-a.s.
    Moreover, $-|X|\leq X_k\leq|X|$ and monotonicity give $\r_{u,t}(|X|)\leq\r_{u,t}(X_k)\leq\r_{u,t}(-|X|)$, so that $|\r_{u,t}(X_k)|\leq|\r_{u,t}(|X|)|+|\r_{u,t}(-|X|)|\in L^p(\F_u)$.
    Hence, again by the Lebesgue property, now at $(s,u)$, $\r_{s,u}(-\r_{u,t}(X_k))\to\r_{s,u}(-\r_{u,t}(X))$ $\P$-a.s., and therefore $\r_{s,t}(X)=\r_{s,u}(-\r_{u,t}(X))$.

    Then, (\textit{a}), (\textit{b}) follow from Theorem~\ref{THM:CS-triad}~(\textit{a}), (\textit{c}) applied to $\tilde\r$.
\end{proof}

    Remarks~\ref{REM:CS-IC-CA} and~\ref{REM:CS-MRG} remain valid on $L^p$, since, by Proposition~\ref{PROP:CS-p-duality}, the minimal penalty of $\r$ on $L^p$ is the penalty $c$ of its restriction to $L^\infty$.

\section{Application to BSDEs and interest rate ambiguity}
\label{sec:bsde}

Backward stochastic differential equations (BSDEs) whose driver is nonincreasing in $y$ induce time-consistent cash-subadditive risk measures, which model ambiguity about interest rates \cite[Section~7]{ElKaroui+Ravanelli_2009_Cash_sub-additive_risk_measures_interest_rate_ambiguity}.
We show that their penalty is the value of a stochastic control problem, in which an ambiguous short rate is optimally scheduled over time, and that \itemref{RM:IC} is its dynamic programming principle.

Throughout, $p=2$, $\mathbb F$ is the augmented filtration of a $d$-dimensional Brownian motion $W$, and $\mathcal P$ is the predictable $\sigma$-algebra on $\Omega\times[0,+\infty)$.
Let $\mu>0$ and let $g:\Omega\times[0,+\infty)\times\R\times\R^d\to\R$ be $\mathcal P\otimes\mathscr B(\R\times\R^d)$-measurable, with $g(\cdot,0,0)=0$, and such that $(y,z)\mapsto g(\omega,v,y,z)$ is convex, nonincreasing in $y$, and $\mu$-Lipschitz.
As usual, we suppress the dependence on $\omega$.
For $0\leq s\leq t$ and $X\in L^2(\F_t)$, let $\r_{s,t}(X):=Y_s$, where $(Y,V)$ is the unique solution of
\[
    Y_v=-X+\int_v^tg(w,Y_w,V_w)\d w-\int_v^tV_w\cdot\!\d W_w,\qquad v\in[0,t],
\]
with $Y$ continuous and adapted, $\E[\sup_{v\leq t}|Y_v|^2]<\infty$, and $V$ predictable, $\E[\int_0^t|V_v|^2\d v]<\infty$ \cite[Theorem~3.1]{Pardoux+Peng_1990_Adapted_solution_backward_stochastic_differential_equation}; see also \cite[Theorem~2.1]{ElKaroui+Peng+Quenez_1997_Backward_stochastic_differential_equations_finance}.

Then $\r$ is a local, convex, cash-subadditive and normalized risk measure on $L^2$, and it is time-consistent (see, e.g., \cite{ElKaroui+Peng+Quenez_1997_Backward_stochastic_differential_equations_finance,ElKaroui+Ravanelli_2009_Cash_sub-additive_risk_measures_interest_rate_ambiguity}; \cite[Proposition~7.3]{ElKaroui+Ravanelli_2009_Cash_sub-additive_risk_measures_interest_rate_ambiguity} treats bounded $X$, and its proof extends to $L^2$ by the comparison theorem \cite[Theorem~2.2]{ElKaroui+Peng+Quenez_1997_Backward_stochastic_differential_equations_finance}). 
By Theorem~\ref{THM:CS-triad-p}, the minimal penalty $c$ of $\r$ satisfies \itemref{RM:IC}, and the essential infimum is attained.

For $(\beta,q)\in\R\times\R^d$, let
\[
    G(v,\beta,q):=\sup_{(y,z)\in\R\times\R^d}\big\{-\beta y+q\cdot z-g(v,y,z)\big\}\in[0,+\infty],
\]
which is $\mathcal P\otimes\mathscr B(\R\times\R^d)$-measurable, since the supremum can be taken over a countable dense set, and nonnegative, since $g(\cdot,0,0)=0$.
Since $g$ is nonincreasing in $y$ and $\mu$-Lipschitz, $G(v,\beta,q)=+\infty$ unless $0\leq\beta\leq\mu$ and $|q|\leq\mu$.
Let $\mathcal U$ be the set of predictable processes $(\beta,q):\Omega\times[0,+\infty)\to [0,\mu]\times\{x\in\R^d:|x|\leq\mu\}$ and, for $(\beta,q)\in\mathcal U$ and $0\leq s\leq t$, let
\begin{align}
    \Lambda^{\beta,q}_{s,t}&:=\exp\bigg(-\int_s^t\beta_v\d v+\int_s^tq_v\cdot\!\d W_v-\frac12\int_s^t|q_v|^2\d v\bigg),
    \\
    c^{\beta,q}_{s,t}&:=\E\bigg[\int_s^t\Lambda^{\beta,q}_{s,v}\,G(v,\beta_v,q_v)\d v\,\bigg|\,\F_s\bigg]\in L^0\big(\F_s;[0,+\infty]\big).
\end{align}
Here $\beta$ is a scenario for the short rate, $q$ is the Girsanov kernel of a scenario for the model, and $G$ penalizes the scenarios according to their plausibility.
Since $q$ is bounded, Novikov's criterion gives $\E\big[\Lambda^{0,q}_{s,t}\big|\F_s\big]=1$, so that $\Lambda^{\beta,q}_{s,t}\in\mathcal Z_{s,t}$ is the conditional density of the probability measure with Girsanov kernel $q$, discounted at the short rate $\beta$.

For $(\beta,q)\in\mathcal U$ and $X\in L^2(\F_t)$, It\^o's formula for $\Lambda^{\beta,q}_{s,\cdot}Y$ on $[s,t]$ gives
\[
    \r_{s,t}(X)=\E\bigg[-\Lambda^{\beta,q}_{s,t}X+\int_s^t\Lambda^{\beta,q}_{s,v}\big(g(v,Y_v,V_v)+\beta_vY_v-q_v\cdot V_v\big)\d v\,\bigg|\,\F_s\bigg],
\]
and $g(v,Y_v,V_v)+\beta_vY_v-q_v\cdot V_v\geq-G(v,\beta_v,q_v)$ by definition of $G$, with equality for a predictable selection of maximizers \cite[Proposition~3.4]{ElKaroui+Peng+Quenez_1997_Backward_stochastic_differential_equations_finance}.
Hence, for all $X\in L^2(\F_t)$,
\begin{equation}
\label{EQ:BSDE-dual}
    \r_{s,t}(X)=\essmax_{(\beta,q)\in\mathcal U}\Big\{\E\bigl[-\Lambda^{\beta,q}_{s,t}X\big|\F_s\bigr]-c^{\beta,q}_{s,t}\Big\};
\end{equation}
see also \cite[Theorem~7.5]{ElKaroui+Ravanelli_2009_Cash_sub-additive_risk_measures_interest_rate_ambiguity} for bounded positions.
Since $\Lambda^{\beta,q}_{s,v}=\Lambda^{\beta,q}_{s,u}\Lambda^{\beta,q}_{u,v}$ for $s\leq u\leq v$, the tower property gives
\begin{equation}
\label{EQ:BSDE-cocycle}
    c^{\beta,q}_{s,t}=c^{\beta,q}_{s,u}+\E\big[\Lambda^{\beta,q}_{s,u}\,c^{\beta,q}_{u,t}\big|\F_s\big],
    \qquad 0\leq s\leq u\leq t.
\end{equation}
As a function of the scenario $(\beta,q)$, the penalty thus satisfies a cocycle property along the factorization $(\Lambda^{\beta,q}_{s,u},\Lambda^{\beta,q}_{u,t})$; see \cite[Remark~4.1]{Drapeau+Kupper+RosazzaGianin+Tangpi_2016_Dual_representation_minimal_supersolutions_convex_BSDEs}, and compare with (Cc) in Remark~\ref{REM:CS-MRG}.
However, a density $Z\in\mathcal Z_{s,t}$ does not determine the scenario: its discount can be accrued along many rate paths $\beta$, which lead to different intermediate values $\Lambda^{\beta,q}_{s,u}$.
The minimal penalty selects the cheapest one.

\begin{proposition}
\label{PROP:BSDE-penalty}
    Let $0\leq s\leq u\leq t$ and $Z\in\mathcal Z_{s,t}$ with $c_{s,t}(Z)<\infty$ $\P$-a.s.
    \begin{enumerate}[label=$(\alph*)$, leftmargin=*]
        \item
            $\ds c_{s,t}(Z)=\essmin\big\{c^{\beta,q}_{s,t}\, :\,  (\beta,q)\in\mathcal U,\ \Lambda^{\beta,q}_{s,t}=Z\big\}$, $\P$-a.s.
        \item
            If $(\beta^*,q^*)$ attains the minimum in $(a)$ and $\Lambda^*:=\Lambda^{\beta^*,q^*}$, then $(\Lambda^*_{s,u},\Lambda^*_{u,t})\in\Phi_u(Z)$ attains the essential infimum in \itemref{RM:IC}, and $c_{s,u}(\Lambda^*_{s,u})=c^{\beta^*,q^*}_{s,u}$, $c_{u,t}(\Lambda^*_{u,t})=c^{\beta^*,q^*}_{u,t}$.
    \end{enumerate}
\end{proposition}

The proof is given in Appendix~\ref{APP:BSDE-penalty}.

\begin{remark}
\label{REM:BSDE-DPP}
    By Proposition~\ref{PROP:BSDE-penalty}, $c_{s,t}(Z)$ is the value of a stochastic control problem (for $s=0$, cf.\ \cite[Remark~3.9]{Drapeau+Kupper+RosazzaGianin+Tangpi_2016_Dual_representation_minimal_supersolutions_convex_BSDEs}), in which the discounted density $\Lambda^{\beta,q}_{s,\cdot}$ is driven from $1$ at time $s$ to $Z$ at time $t$ at minimal expected cost.
    Fixing the state $Z^1=\Lambda^{\beta,q}_{s,u}$ at the intermediate date splits this problem into two problems on $[s,u]$ and $[u,t]$, with values $c_{s,u}(Z^1)$ and $\E[Z^1c_{u,t}(Z^2)|\F_s]$, $Z^2=Z/Z^1$, and \itemref{RM:IC} is the dynamic programming principle.

    The relation between time-consistency and Bellman's principle is well known on the primal side, where time-consistency of $\r$ itself is read as a dynamic programming principle \cite{Artzner+Delbaen+Eber+Heath+Ku_2007_Coherent_multiperiod_risk_adjusted_values_Bellmans_principle,Follmer+Penner_2006_Convex_risk_measures_dynamics_their_penalty_functions,Bielecki+Cialenco+Pitera_2017_Survey_time_consistency_dynamic_risk_measures}.
    To our knowledge, the corresponding dual statement, with the penalty as the value function and the factorizations as intermediate states, has not been made explicit.
\end{remark}

\begin{remark}
\label{REM:BSDE-CA}
    If $g$ does not depend on $y$, then $\r$ is cash-additive, $G(v,\beta,q)=+\infty$ for $\beta\neq0$, the kernel $q$ is determined by $Z=\Lambda^{0,q}_{s,t}$, and Proposition~\ref{PROP:BSDE-penalty}~(\textit{a}) reduces to the representation $c_{s,t}(Z)=\E\big[Z\int_s^tG(v,0,q_v)\d v\big|\F_s\big]$ of \cite[Theorem~3.2~(i)]{Delbaen+Peng+RosazzaGianin_2010_Representation_penalty_term_dynamic_concave_utilities}, with no minimization.
\end{remark}

\begin{example}[Ambiguous interest rates]
\label{EX:BSDE-corridor}
    The basic model of interest rate ambiguity in \cite{ElKaroui+Ravanelli_2009_Cash_sub-additive_risk_measures_interest_rate_ambiguity} is that of a regulator who only knows that the short rate lies in a corridor $[r,R]$, e.g., between a deposit and a lending rate, and who evaluates positions under the worst rate.
    We show that, in this case, \itemref{RM:IC} states that the admissible scenarios can be switched at any intermediate date, and that the restriction factorization is not optimal as soon as $r>0$.
    
    Let $0\leq r\leq R$, $\nu\geq0$ and $g(v,y,z):=Ry^--ry^++\nu|z|$, which is $\mu$-Lipschitz with $\mu:=R\vee\nu$ and, for $\nu=0$, is the driver of \cite[Example~7.2]{ElKaroui+Ravanelli_2009_Cash_sub-additive_risk_measures_interest_rate_ambiguity}; the term $\nu|z|$ adds ambiguity about the model.
    Then $G(v,\beta,q)=0$ if $r\leq\beta\leq R$ and $|q|\leq\nu$, and $G(v,\beta,q)=+\infty$ otherwise, so that $\r_{s,t}(X)=\essmax_{Z\in\mathcal D_{s,t}}\E[-ZX|\F_s]$ is the worst-case discounted expected loss over
    \[
        \mathcal D_{s,t}:=\big\{\Lambda^{\beta,q}_{s,t}\, :\,  (\beta,q):\Omega\times[0,+\infty)\to\R\times\R^d\text{ predictable},\ r\leq\beta\leq R,\ |q|\leq\nu\big\},
    \]
    i.e., over the short rates in the corridor $[r,R]$ and the models with kernel bounded by $\nu$.
    
    By \eqref{EQ:BSDE-dual} and Proposition~\ref{PROP:BSDE-penalty}, $c_{s,t}(Z)=0$ for $Z\in\mathcal D_{s,t}$ and $\E[c_{s,t}(Z)]=+\infty$ otherwise, and \itemref{RM:IC} reduces to the stability under products 
    \[
        \mathcal D_{s,t}=\{Z^1Z^2:Z^1\in\mathcal D_{s,u},\ Z^2\in\mathcal D_{u,t}\}:
    \]
    following an admissible scenario up to time $u$ and another one afterwards yields an admissible scenario.
    For $Z\in\mathcal D_{s,t}$, every factorization in $\mathcal D_{s,u}\times\mathcal D_{u,t}$ is optimal.
    If $r>0$ and $u<t$, the restriction factorization is not: $\E[Z\oslash Z_u|\F_u]=1$, whereas $\E[Z^2|\F_u]\leq e^{-r(t-u)}<1$ for $Z^2\in\mathcal D_{u,t}$, so that $c_{u,t}(Z\oslash Z_u)=+\infty$.
\end{example}

\begin{example}[Penalized interest rates]
\label{EX:BSDE-quadratic}
    In Example~\ref{EX:BSDE-corridor}, all rates in the corridor are equally plausible.
    A smoother description of ambiguity, as in the multiplier and variational preferences of \cite{Hansen+Sargent_2001_Robust_Control_Model_Uncertainty,Maccheroni+Marinacci+Rustichini_2006_Ambiguity_aversion_robustness__variational_representation_preferences}, admits rates above a reference rate but penalizes them according to their distance from it.
    Then the penalty is not trivial, and \itemref{RM:IC} is a genuine optimization, which we solve explicitly.

    Let $\theta,R>0$ and $g(v,y,z):=\max_{0\leq\beta\leq R}\{-\beta y-\beta^2/(2\theta)\}$, which is $\mu$-Lipschitz with $\mu:=R$.
    Then $G(v,\beta,q)=\beta^2/(2\theta)$ if $0\leq\beta\leq R$ and $q=0$, and $G(v,\beta,q)=+\infty$ otherwise: the reference rate is $0$, rates up to $R$ are admitted, and $\theta$ measures the tolerance towards high rates.
    Fix $s<u<t$ and a discount factor $z\in L^\infty(\F_s)$ known at time $s$, with $0<z\leq1$.
    We compute $c_{s,t}(z)$ by Proposition~\ref{PROP:BSDE-penalty}~(\textit{a}), i.e., by minimizing $c^{\beta,q}_{s,t}$ over the scenarios $(\beta,q)\in\mathcal U$ with $\Lambda^{\beta,q}_{s,t}=z$.
    
    Since $G=+\infty$ for $q\neq0$, a scenario has finite cost only if $q=0$ on $[s,t]$; then $\Lambda_v:=\Lambda^{\beta,0}_{s,v}=\exp\big(-\int_s^v\beta_w\d w\big)$ is absolutely continuous and nonincreasing in $v$, with $\Lambda_s=1$, $\Lambda_t=z$ and $\beta_v=-\Lambda'_v/\Lambda_v$.
    Then the running cost is $\Lambda_vG(v,\beta_v,0)=\Lambda_v\beta_v^2/(2\theta)=(\Lambda'_v)^2/(2\theta \Lambda_v)=\frac2\theta\big((\sqrt{\Lambda_v})'\big)^2$, so that
    \[
        c^{\beta,0}_{s,t}=\frac2\theta\,\E\bigg[\int_s^t\Big(\big({\sqrt{\Lambda_v}}\big)'\Big)^2\d v\,\bigg|\,\F_s\bigg].
    \]
    Along each path, $\sqrt \Lambda$ goes from $1$ to $\sqrt z$, and the Cauchy--Schwarz inequality gives
    \[
        \big(1-\sqrt z\big)^2=\bigg(\int_s^t\big(\sqrt{\Lambda_v}\big)'\d v\bigg)^2\leq(t-s)\int_s^t\Big(\big(\sqrt{\Lambda_v}\big)'\Big)^2\d v,
    \]
    with equality if and only if $(\sqrt{\Lambda_v})'$ is constant, i.e., $\sqrt{\Lambda_v}=\sqrt{\Lambda^*_v}:=1-a(v-s)$ with $a:=(1-\sqrt z)/(t-s)$.
    Hence $c^{\beta,0}_{s,t}\geq2(1-\sqrt z)^2/(\theta(t-s))$, with equality if and only if $\Lambda=\Lambda^*$.
    The path $\Lambda^*$ corresponds to the rate $\beta^*_v=-(\Lambda^*_v)'/\Lambda^*_v=2a/(1-a(v-s))$, which is $\F_s$-measurable and nondecreasing in $v$, so that $0\leq\beta^*_v\leq\beta^*_t=2a/\sqrt z$.
    Hence $(\beta^*,0)\in\mathcal U$ if and only if $2a/\sqrt z\leq R$, i.e.,
    \begin{equation}
    \label{EQ:BSDE-quad-z}
        z\geq\bigg(1+\frac{R(t-s)}2\bigg)^{-2},
    \end{equation}
    which we assume from now on: the discount to be generated is not so large that the optimal rate exceeds the bound $R$.
    Then the minimum is attained at $(\beta^*,0)$.

    By Proposition~\ref{PROP:BSDE-penalty}, it follows that
    \[
        c_{s,t}(z)=\frac{2(1-\sqrt z)^2}{\theta(t-s)},
        \qquad
        \sqrt{Z^1}=\sqrt{\Lambda^*_u}=\frac{t-u}{t-s}+\frac{u-s}{t-s}\sqrt z,
    \]
    where $(Z^1,z/Z^1)$ is the unique optimal factorization of $z$ at time $u$, since every optimal factorization is generated by a minimizing scenario, by Proposition~\ref{PROP:BSDE-penalty}~(\textit{a}) and \eqref{EQ:BSDE-cocycle}.

    The optimal factorization describes how the total discount $z$ is split between the two periods: $Z^1=\Lambda^*_u$ is the discount accrued on $[s,u]$, and $z/Z^1$ the one accrued on $[u,t]$.
    Two allocations are natural candidates: the restriction factorization $Z^1=z$, which accrues the whole discount on $[s,u]$, and the constant-rate allocation $Z^1=z^{(u-s)/(t-s)}$, which accrues it proportionally to time.
    The optimal one is neither: if $z<1$, the weighted arithmetic--geometric mean inequality gives
    \[
        z<z^{(u-s)/(t-s)}<\Lambda^*_u<1,
    \]
    i.e., the optimal factorization accrues on $[s,u]$ less discount than both candidates, and postpones the rest to $[u,t]$.
\end{example}

\appendix

\section{\texorpdfstring{Proof of Proposition~\ref{PROP:CS-p-duality}}{Proof of the Lp dual representation}}
\label{APP:CS-p-duality}

\begin{proof}[Proof of Proposition~\ref{PROP:CS-p-duality}]
    Fix $Z\in\mathcal Z_{s,t}$ and $X\in L^p(\F_t)$.
    \smallskip

    \emph{Step 1: integrability.}
    We first show that, for all $U\in L^p_+(\F_t)$ and $\l>0$,
    \begin{equation}
    \label{EQ:CS-p-integrability}
        \E[ZU|\F_s]\leq\frac1\l\Big(c_{s,t}(Z)+\r_{s,t}(-\l U)\Big).
    \end{equation}
    For $n\in\N$, $-\l(U\wedge n)\in L^\infty(\F_t)$ and $-\l(U\wedge n)\geq-\l U$, so that \eqref{EQ:CS-penalty} and monotonicity give $\l\,\E[Z(U\wedge n)|\F_s]\leq c_{s,t}(Z)+\r_{s,t}(-\l(U\wedge n))\leq c_{s,t}(Z)+\r_{s,t}(-\l U)$.
    Monotone convergence gives \eqref{EQ:CS-p-integrability}, and the first claim follows with $U:=|X|$ and $\l:=1$, since $\r_{s,t}(-|X|)$ is finite.
    \smallskip

    \emph{Step 2: the inequality ``$\geq$'' in \eqref{EQ:CS-representation}.}
    Write $h_{Z'}(X):=\E[-Z'X|\F_s]-c_{s,t}(Z')$ for $Z'\in\mathcal Z_{s,t}$, and $X_k:=(X\wedge k)\vee(-k)$, $k\in\N$.
    Since $X_k\in L^\infty(\F_t)$, \eqref{EQ:CS-penalty} gives $h_Z(X_k)\leq\r_{s,t}(X_k)$.
        As $k\to\infty$, $\r_{s,t}(X_k)\to\r_{s,t}(X)$ by Lemma~\ref{LEM:CS-p-basic}~(\textit{a}).
    Moreover, $ZX_k\to ZX$ with $|ZX_k|\leq Z|X|$, where $Z|X|$ need not be integrable; however, on each $B_n:=\{\E[Z|X||\F_s]\leq n\}\in\F_s$ the variable $\1_{B_n}Z|X|$ is integrable, so that conditional dominated convergence gives $h_Z(X_k)\to h_Z(X)$ on $B_n$.
    Since $\bigcup_nB_n\supseteq\{c_{s,t}(Z)<\infty\}$ by Step~1, $h_Z(X_k)\to h_Z(X)$ on $\{c_{s,t}(Z)<\infty\}$, while $h_Z(X)=-\infty$ on its complement.
    Hence $h_Z(X)\leq\r_{s,t}(X)$, and, since $Z$ is arbitrary, $R(X):=\esssup_{Z'\in\mathcal Z_{s,t}}h_{Z'}(X)\leq\r_{s,t}(X)$.
    \smallskip

    \emph{Step 3: the inequality ``$\leq$'' in \eqref{EQ:CS-representation}.}
    By Lemma~\ref{LEM:CS-p-basic}~(\textit{b}), Proposition~\ref{PROP:CS-duality} applies to $\tilde\r$, and $c_{s,t}$ is the minimal penalty of $\tilde\r_{s,t}$.
    Fix $k\in\N$.
    By Proposition~\ref{PROP:CS-duality}~(\textit{c}), there are $Z^n\in\mathcal Z_{s,t}$ with $-k-1\leq h_{Z^n}(X_k)\uparrow\r_{s,t}(X_k)$, so that, by monotone convergence, $\E[h_{Z^n}(X_k)]\to\E[\r_{s,t}(X_k)]$.
    Choosing $Z_k:=Z^n$ for $n$ large enough, we obtain $\E[h_{Z_k}(X_k)]\geq\E[\r_{s,t}(X_k)]-\frac1k$.
    Since $h_{Z_k}(X_k)\geq-k-1$, $c_{s,t}(Z_k)$ is finite and integrable, and since $Z_kX_k$ is integrable, the tower property gives
    \begin{equation}
    \label{EQ:Zk}
        \E[-Z_kX_k]-\E[c_{s,t}(Z_k)]=\E[h_{Z_k}(X_k)]\geq\E[\r_{s,t}(X_k)]-\frac1k.
    \end{equation}

    We now bound $\E[c_{s,t}(Z_k)]$ uniformly in $k$.
    Taking expectations in \eqref{EQ:CS-p-integrability} with $Z=Z_k$, $U=|X|$ and $\l=2$ gives
    \begin{equation}
    \label{EQ:CS-p-ZkX}
        \E[Z_k|X|]\leq\frac12\E[c_{s,t}(Z_k)]+\frac12\E[\r_{s,t}(-2|X|)].
    \end{equation}
    Since $-Z_kX_k\leq Z_k|X_k|\leq Z_k|X|$ and, by monotonicity, $\r_{s,t}(X_k)\geq\r_{s,t}(|X|)$ (as $X_k\leq|X|$), \eqref{EQ:Zk} and \eqref{EQ:CS-p-ZkX} yield
    \begin{align}
        \E[c_{s,t}(Z_k)]
        &\leq\E[Z_k|X|]-\E[\r_{s,t}(X_k)]+\frac1k\\
        &\leq\frac12\E[c_{s,t}(Z_k)]+\frac12\E[\r_{s,t}(-2|X|)]-\E[\r_{s,t}(|X|)]+1.
    \end{align}
    As $\E[c_{s,t}(Z_k)]<\infty$, rearranging yields
    \[
        \E[c_{s,t}(Z_k)]\leq\kappa:=\E[\r_{s,t}(-2|X|)]-2\,\E[\r_{s,t}(|X|)]+2,
    \]
    where $\kappa<\infty$ because $\r_{s,t}(-2|X|),\r_{s,t}(|X|)\in L^p(\F_s)\subseteq L^1(\F_s)$.
    In particular, $Z_k|X|$ is integrable by \eqref{EQ:CS-p-ZkX}.

    We now pass from $X_k$ to $X$.
    Since $c_{s,t}(Z_k)<\infty$ $\P$-a.s.\ and $Z_kX$, $Z_kX_k$ are integrable,
    \[
        h_{Z_k}(X)=h_{Z_k}(X_k)-\E[Z_k(X-X_k)|\F_s]\geq h_{Z_k}(X_k)-\E[Z_k|X-X_k||\F_s].
    \]
    Taking expectations and using \eqref{EQ:Zk}, together with \eqref{EQ:CS-p-integrability} with $Z=Z_k$ and $U:=|X-X_k|=(|X|-k)^+$, we obtain for every $\l>0$
    \[
        \E[h_{Z_k}(X)]
        \geq\E[\r_{s,t}(X_k)]-\frac1k-\frac1\l\Big(\kappa+\E\big[\r_{s,t}\big(-\l(|X|-k)^+\big)\big]\Big).
    \]
    As $k\to\infty$, $\E[\r_{s,t}(X_k)]\to\E[\r_{s,t}(X)]$ and $\E[\r_{s,t}(-\l(|X|-k)^+)]\to0$.
    Indeed, by the Lebesgue property in Lemma~\ref{LEM:CS-p-basic}~(\textit{a}), with $U:=|X|$ and $U:=\l|X|$, respectively, we have $\r_{s,t}(X_k)\to\r_{s,t}(X)$ and $\r_{s,t}(-\l(|X|-k)^+)\to\r_{s,t}(0)=0$ $\P$-a.s.
    Moreover, by monotonicity,
    \[
        \r_{s,t}(|X|)\leq\r_{s,t}(X_k)\leq\r_{s,t}(-|X|),
        \qquad
        0=\r_{s,t}(0)\leq\r_{s,t}\big(-\l(|X|-k)^+\big)\leq\r_{s,t}(-\l|X|),
    \]
    with integrable bounds, so that dominated convergence applies.
    Since $h_{Z_k}(X)\leq R(X)\leq\r_{s,t}(X)$ by Step~2, we obtain, for every $\l>0$,
    \[
        \E[\r_{s,t}(X)]-\frac\kappa\l
        \leq\liminf_k\E[h_{Z_k}(X)]
        \leq\E[R(X)]
        \leq\E[\r_{s,t}(X)].
    \]
    Letting $\l\to\infty$ gives $\E[R(X)]=\E[\r_{s,t}(X)]$.
    As $R(X)\leq\r_{s,t}(X)$ and $\r_{s,t}(X)$ is integrable, $R(X)=\r_{s,t}(X)$ $\P$-a.s., which is \eqref{EQ:CS-representation}.
    \smallskip

    \emph{Step 4: minimality.}
    If \eqref{EQ:CS-representation} holds on $L^p(\F_t)$ with $\a$, then it holds on $L^\infty(\F_t)$, and $c_{s,t}\leq\a$ by Proposition~\ref{PROP:CS-duality}~(\textit{b}) applied to $\tilde\r$.
\end{proof}

\section{\texorpdfstring{Proof of Proposition~\ref{PROP:BSDE-penalty}}{Proof of the representation of the penalty for BSDEs}}
\label{APP:BSDE-penalty}

\begin{proof}[Proof of Proposition~\ref{PROP:BSDE-penalty}]
    \emph{Step 1.}
    For $(\beta,q)\in\mathcal U$, the dual representation \eqref{EQ:BSDE-dual} gives $c^{\beta,q}_{s,t}\geq\E[-\Lambda^{\beta,q}_{s,t}X|\F_s]-\r_{s,t}(X)$ for all $X\in L^\infty(\F_t)$.
    Hence, by definition \eqref{EQ:CS-penalty} of the penalty,
    \begin{equation}
    \label{EQ:BSDE-min}
        c_{s,t}\big(\Lambda^{\beta,q}_{s,t}\big)\leq c^{\beta,q}_{s,t}.
    \end{equation}

    \emph{Step 2 (linearization).}
    For $(\beta,q)\in\mathcal U$, set $\Lambda_v:=\Lambda^{\beta,q}_{s,v}$, $b:=\Lambda\beta$ and $\eta:=\Lambda q$ on $[s,t]$, so that
    \begin{equation}
    \label{EQ:BSDE-linear}
        \Lambda_v=1-\int_s^vb_w\d w+\int_s^v\eta_w\cdot\!\d W_w,
        \qquad
        0\leq b\leq\mu\Lambda,\quad|\eta|\leq\mu\Lambda.
    \end{equation}
    Since $\Lambda_v>0$, $\Lambda_vG(v,\beta_v,q_v)=\hat G(v,\Lambda_v,b_v,\eta_v)$, where
    \[
        \hat G(v,\lambda,b,\eta):=\sup_{(y,z)\in\R\times\R^d}\big\{-by+\eta\cdot z-\lambda g(v,y,z)\big\}\geq0
    \]
    is convex and lower semicontinuous in $(\lambda,b,\eta)\in[0,+\infty)\times\R\times\R^d$, as a supremum of linear functions.
    Conversely, let $\mathcal K$ be the set of predictable $(b,\eta)\in L^2(\Omega\times[s,t])$ satisfying the inequalities in \eqref{EQ:BSDE-linear}, with $\Lambda$ defined by the identity in \eqref{EQ:BSDE-linear}.
    For $(b,\eta)\in\mathcal K$, the processes $\beta:=(b/\Lambda)\1_{\{\Lambda>0\}}$ and $q:=(\eta/\Lambda)\1_{\{\Lambda>0\}}$ belong to $\mathcal U$, and $\Lambda=\Lambda^{\beta,q}_{s,\cdot}$, since $b=\eta=0$ on $\{\Lambda=0\}$ and both processes solve $\d\Lambda=\Lambda(-\beta\d v+q\cdot\!\d W)$, $\Lambda_s=1$.
    Hence, for $Z\in L^1(\F_t)$,
    \[
        \phi(Z):=\inf\big\{\E[c^{\beta,q}_{s,t}]:(\beta,q)\in\mathcal U,\ \Lambda^{\beta,q}_{s,t}=Z\big\}
        =\inf\big\{J(b,\eta):(b,\eta)\in\mathcal K,\ \Lambda_t=Z\big\},
    \]
    with $\inf\varnothing:=+\infty$ and $J(b,\eta):=\E\big[\int_s^t\hat G(v,\Lambda_v,b_v,\eta_v)\d v\big]$.

    \emph{Step 3 (compactness).}
    The set $\mathcal K$ is convex, closed and bounded in $L^2$, since $\E[\Lambda_v^2]\leq e^{\mu^2(v-s)}$, hence weakly compact.
    The map $(b,\eta)\mapsto\Lambda_t$ is affine and continuous from $L^2$ to $L^2(\F_t)$, hence weakly continuous.
    The functional $J$ is convex and, by Fatou's lemma, lower semicontinuous on $\mathcal K$, hence weakly lower semicontinuous.
    Since $J$ is convex, $\mathcal K$ is convex and $(b,\eta)\mapsto\Lambda_t$ is affine, the function $((b,\eta),Z)\mapsto J(b,\eta)$ on $\{(b,\eta)\in\mathcal K,\ \Lambda_t=Z\}$, and $+\infty$ elsewhere, is jointly convex, so that $\phi$ is convex by \cite[Theorem~1]{Rockafellar_1974_Conjugate_duality_optimization}.
    It is also lower semicontinuous on $L^1(\F_t)$: let $Z_n\to Z$ in $L^1(\F_t)$ and, passing to a subsequence, assume $\phi(Z_n)\to\liminf_n\phi(Z_n)<\infty$; choose $(b_n,\eta_n)\in\mathcal K$, for $n$ large enough, with $\Lambda^n_t=Z_n$ and $J(b_n,\eta_n)\leq\phi(Z_n)+1/n$; a further subsequence converges weakly to some $(b,\eta)\in\mathcal K$, with $\Lambda_t=Z$ and $J(b,\eta)\leq\lim_n\phi(Z_n)$.
    Taking $Z_n:=Z$, the infimum defining $\phi(Z)$ is attained whenever it is finite.

    \emph{Step 4 (duality).}
    By \eqref{EQ:BSDE-dual} with $X=0$, there is $(\beta,q)\in\mathcal U$ with $c^{\beta,q}_{s,t}=0$, so that $\phi$ is proper.
    For $X\in L^\infty(\F_t)$, the essential supremum in \eqref{EQ:BSDE-dual} is attained, so that
    \[
        \sup_{Z\in L^1(\F_t)}\big\{\E[-ZX]-\phi(Z)\big\}
        =\sup_{(\beta,q)\in\mathcal U}\E\big[-\Lambda^{\beta,q}_{s,t}X-c^{\beta,q}_{s,t}\big]
        =\E[\r_{s,t}(X)].
    \]
    By Step~3 and the Fenchel--Moreau theorem \cite[Theorem~A.65]{Follmer+Schied_2016_Stochastic_finance} on $L^1(\F_t)$, whose dual is $L^\infty(\F_t)$, and by \eqref{EQ:CS-expectation_c}, for all $Z\in\mathcal Z_{s,t}$,
    \[
        \phi(Z)=\sup_{X\in L^\infty(\F_t)}\E\bigl[-ZX-\r_{s,t}(X)\bigr]=\E[c_{s,t}(Z)].
    \]

(\textit{a}).
    Assume first $\E[c_{s,t}(Z)]<\infty$.
    Since $\phi(Z)=\E[c_{s,t}(Z)]$, by Step~3 there is $(\beta^*,q^*)\in\mathcal U$ with $\Lambda^{\beta^*,q^*}_{s,t}=Z$ and $\E[c^{\beta^*,q^*}_{s,t}]=\E[c_{s,t}(Z)]$.
    By \eqref{EQ:BSDE-min}, $c_{s,t}(Z)\leq c^{\beta,q}_{s,t}$ for every $(\beta,q)\in\mathcal U$ with $\Lambda^{\beta,q}_{s,t}=Z$; hence $c^{\beta^*,q^*}_{s,t}=c_{s,t}(Z)$.
    In general, let $A_n:=\{n-1\leq c_{s,t}(Z)<n\}\in\F_s$, let $(\beta^0,q^0)\in\mathcal U$ with $c^{\beta^0,q^0}_{s,t}=0$ (Step~4), so that $c_{s,t}\big(\Lambda^{\beta^0,q^0}_{s,t}\big)=0$ by \eqref{EQ:BSDE-min}, and let $Z_n:=\1_{A_n}Z+\1_{A_n^c}\Lambda^{\beta^0,q^0}_{s,t}$.
    By Proposition~\ref{PROP:CS-duality}~(\textit{a}), $c_{s,t}(Z_n)=\1_{A_n}c_{s,t}(Z)\leq n$, so that the first case gives $(\beta^n,q^n)\in\mathcal U$ with $\Lambda^{\beta^n,q^n}_{s,t}=Z_n$ and $c^{\beta^n,q^n}_{s,t}=c_{s,t}(Z_n)$.
    Since $(A_n)_n$ is an $\F_s$-partition of $\Omega$, $(\beta^*,q^*):=\sum_n\1_{A_n}(\beta^n,q^n)$ on $(s,t]$ belongs to $\mathcal U$, and $\Lambda^{\beta^*,q^*}_{s,t}=\sum_n\1_{A_n}Z_n=Z$ and $c^{\beta^*,q^*}_{s,t}=\sum_n\1_{A_n}c_{s,t}(Z_n)=c_{s,t}(Z)$; minimality follows from \eqref{EQ:BSDE-min} as before.

(\textit{b}).
    By (\textit{a}), by \eqref{EQ:BSDE-cocycle}, by \eqref{EQ:BSDE-min} at $(s,u)$ and at $(u,t)$, and by \eqref{EQ:CS-subcocycle}, letting $c^{\beta^*,q^*}=:c^*$,
    \begin{align}
        c_{s,t}(Z)
        =c^*_{s,u}+\E\big[\Lambda^*_{s,u}\,c^*_{u,t}\big|\F_s\big]
        &\geq c_{s,u}(\Lambda^*_{s,u})+\E\big[\Lambda^*_{s,u}\,c_{u,t}(\Lambda^*_{u,t})\big|\F_s\big]
        \\
        &=\Gamma_{s,u,t}(\Lambda^*_{s,u},\Lambda^*_{u,t})
        \geq c_{s,t}(Z).
    \end{align}
    Hence equality holds throughout, and $(\Lambda^*_{s,u},\Lambda^*_{u,t})\in\Phi_u(Z)$ attains the essential infimum in \itemref{RM:IC}.
    Since $c_{s,t}(Z)<\infty$ and both inequalities in \eqref{EQ:BSDE-min} hold termwise, $c_{s,u}(\Lambda^*_{s,u})=c^*_{s,u}$ and $\E\big[\Lambda^*_{s,u}\big(c^*_{u,t}-c_{u,t}(\Lambda^*_{u,t})\big)\big|\F_s\big]=0$, where the integrand is nonnegative; since $\Lambda^*_{s,u}>0$, $c_{u,t}(\Lambda^*_{u,t})=c^*_{u,t}$.
\end{proof}

\printbibliography

\end{document}